## Highlights

**Unbiased higher-order frictional contact using midplane and patch based segment-to-segment penalty method**

Indrajeet Sahu  , Nik Petrinic

- High-accuracy unbiased frictional contact formulation is presented for higher-order finite elements
- Higher-order elements provide smoother contact pressure in static contacts and more accurate momentum exchange in dynamic contacts
- The unbiased nature of the formulation aids in the smooth handling of frictional self-contact problems and large deformation problems
- The provided formulation can represent the nonlinear distribution of frictional forces on contacting nodes

# Unbiased higher-order frictional contact using midplane and patch based segment-to-segment penalty method

Indrajeet Sahu [a,*], Nik Petrinic [a]

[a]*Department of Engineering Science, University of Oxford, Parks Road, Oxford, OX1 3PJ, Oxfordshire, United Kingdom*

**Abstract**

A highly accurate, single-pass, unbiased frictional contact algorithm for higher-order elements based on the concept of midplane is presented. Higher-order elements offer a lucrative choice for contact problems as they can better represent the curvature of original geometries compared to the first-order elements. Compressive and frictional contact constraints are applied over the contact pairs of sub-segments obtained by the subdivision of higher-order segments. The normal traction depends upon the penalisation of true interpenetration, and frictional traction depends upon relative sliding between sub-segments over their shared patches. The midplane constructed by linearised subfacets can be corrected to account for local curvature of interacting physical surfaces. Demonstrated through multiple tests, the use of higher-order elements surpasses the accuracy of first-order elements for curved geometries. Its versatility extends from static to dynamic conditions for flat and curved interfaces including frictional contact. The presented examples include contact patch test, Hertzian contact, elastic collision, rotation of concentric surfaces, frictional sliding, self-contact and inelastic collision problems. Here, contact patch test matches the accuracy of finite elements and Hertzian contact shows smoother solution compared to first-order meshes. The elastic collision problem highlights the utility of the algorithm in accurate prediction of configuration changes in multibody systems. The frictional sliding demonstrates the ability to represent the expected nonlinear distribution of nodal forces for the higher-order elements. The large deformation problems, e.g. self-contact and inelastic collision, specifically benefit from the accuracy in surface representation using higher-order discretisation and continuous contact constraint imposition on such surfaces during deformation.

## 1. Introduction

Finite element solution of frictional contact among complex surfaces in multibody sytems needs special attention. Commonly used first-order elements can not adequately represent the curvature of complex surfaces. However, higher-order elements represent the curvature better due to their nonlinear shape functions. They are also effective in modelling flexure without reduced quadrature and hourglass control, as well as in the issues of volumetric locking in unstructured meshes [1]. Contact constraint enforcement techniques need to account for complexities associated with higher-order basis functions for both compressive and frictional loads. Although finite element solutions using first-order elements together with h-refinement can generate practical solutions, the recently increased development on p-refinement, in particular, second-order elements, provides an active opportunity for techniques that can offer higher accuracy in solutions [2][3][4]. Additionally, the higher-order elements can often achieve the same level of accuracy as linear ones with fewer elements, which has the potential for better computational efficiency. The higher-order elements also have a better convergence rate with the refinement of meshes. In light of this, a contact interaction methodology accounting for the higher-order interpolation of geometry and solution on the surface nodes is presented in this work.

Typically, the general architecture of solving contact problems with finite element methods involves two aspects - contact discretisation and constraint enforcement. When two bodies composed of finite elements interact, pairs of discrete entities are formed on the surface to enforce contact constraints over them. The Node-to-segment (NTS) based discretisation [5][6] is the most widely used technique, wherein pairs between nodes on one surface and the segments on opposite surfaces are formed to define contact interaction and impose the contact constraint disallowing nodes on slave surface from penetrating the master surface. Due to its limited accuracy and inability to transfer even a uniform pressure in contact patch test [7], segment-to-segment (STS) methods are more appropriate for problems where contact traction accuracy is critical. In STS methods, constraints are applied on pairs of segments (or facets) on opposite discretised surfaces in a continuous manner [8], and this is a physically more realistic approach, as is evident in its robustness and accuracy. Constraint enforcement generally employs either penalty or Lagrange multiplier-based methods. While Lagrange multiplier methods enforce the impenetrability condition exactly compared to the penalty method, it incurs an additional cost due to increased system size, whereas the penalty method obviates the need for additional equations and is computationally more efficient, thus being the preferred choice in the present work.

In the last decade and a half, a significant interest has developed in formulating methods for contact between higher-order elements for both implicit and explicit finite element methods. In implicit methods, mortar methods initially used for domain decomposition [9] have been extensively studied for STS interaction by having enforcement of non-penetration condition in a weak (weighted) sense instead of the strong imposition on the entire surface, thereby having the continuity of traction through a weak formulation. In this approach, the traction is defined on one side chosen as mortar by using Lagrange multiplier basis functions together with the discrete nodal Lagrange multipliers [10, 2],[11]. It has been used in both 2D frictionless contact [12] and frictional cases [13][14], as well as in 3D problems [15][16][17] with the dual-mortar formulation for linear[18][19] and quadratic shape functions [3][20]. The dual-mortar methods employ a biorthogonality relationship between Lagrange multipliers and dual-shape functions, such that these methods are able to condense Lagrange multipliers to avoid large system sizes. The mortar methods, however, have a natural bias in the definition of contact with the choice of mortar and non-mortar sides. Some other works on second-order elements can be referred to in [21] for smoothing procedure with multi-point constraints, [22] for covariant contact description, and [23] for 2D frictional contact using splines and Hermite cubic interpolation.

Second-order elements have also been recently extensively studied for their performance in explicit methods for different numbers of nodes and integration rules [24] [25] and for a contact using double pass NTS scheme [4] where the solution is the weighted average of both passes. However, the use of the NTS scheme with complex higher-order elements significantly affects the accuracy of the solution, and the imposition of contact constraints on higher-order elements is a complex challenge with the definition of contact region and enforcement of constraints. Although, in general, the use of explicit methods in common high-speed impact cases might not require high accuracy of surface traction, its use in slower applications where it can potentially compete with implicit methods would depend upon the accuracy of contact detection and contact traction. This accuracy is better achievable using surface interaction-based methods where the contact constraint is imposed continuously over the entire surface represented by complex higher-order basis functions. Additionally, applications involving multiple successive contacts transferring momentum will also be majorly affected by the direction of motion imparted to different bodies.

In the recent STS based unbiased works [26][27], contact traction is resolved in an unbiased manner using the midplane between interacting facets by employing a uniformly imposed penalty. This methodology, initially proposed for facets represented by linear shape functions, is being extended here to contact between facets having higher-order basis functions and stronger curvature. The single-pass unbiased nature of the midplane-based method is preserved here, thereby avoiding the weighted dual pass route of eliminating bias in contact definition generally used in most NTS and STS methods. In this paper, contact between higher-order elements is primarily studied through contact between 27-noded second-order Lagrangian elements; however, a conceptual idea for similar implementation using a 20-noded serendipity element is also provided. Focusing on the second-order Lagrangian finite element, a facet-to-facet (or segment-to-segment) pair refers to the contact between two 9-noded facets, and a subfacet-to-subfacet (or subsegment-to-subsegment) pair refers to the contact between 4-noded subfacets decomposed from the main 9-noded facet as is typical in the literature. This work utilises an explicit finite element method with HRZ lumping of mass matrix and solved using a central difference scheme.

This paper is organised as follows- Section 2 describes the strong form of contact problem between two solids and its development into a weak form for formulating the finite element-based problem. Section 3 introduces and explains the unbiased frictional contact methodology by focusing on second-order Lagrangian elements and brings attention to a source of error due to the linear approximation of the subfacets which is inherent in the typical facet subdivision strategies. A pseudo-code for algorithmic implementation of the methodology is presented in Section 4 before proposing an idea for facet division in serendipity elements in 5. Numerous examples are presented in Section 6 to test the robustness of the proposed method for both frictionless and frictional contact. Finally, a conclusion to the paper is presented in Section 7.

## 2. Problem description

For contact between any two solids $\Omega^i$ ($i = 1, 2$) in $\mathbb{R}^3$, their reference and current configurations can be denoted using any spatial point in the domain as $X^i$ and $x^i$ connected by its displacement $u^i(X, t)$, i.e., $x^i = X^i + u^i$. The force-momentum equation representing the boundary value problem in the strong form at any point can be written as

$$\begin{aligned}
\nabla \cdot \sigma^i + b^i &= \rho^i \partial_t^2 x & \forall \, x^i \in \Omega^i \times (0, T) \\
u^i(x^i) &= \overline{u}^i & \forall \, x^i \in \gamma_u^i \times (0, T) \\
\sigma^i n^i &= \overline{t}_\sigma^i & \forall \, x^i \in \gamma_\sigma^i \times (0, T)
\end{aligned} \quad (1)$$

where the symbols $\sigma^i$, $b^i$ and $\rho^i$ represent Cauchy stress, body force per unit volume and density at any point $x^i$ in $\Omega^i$. While the first equation is a result of the conservation of momentum, the second and third equations impose Dirichlet and Neumann boundary conditions on the $\gamma_u^i$ and $\gamma_\sigma^i$, respectively. The contact zone formed between the two bodies during the motion and deformation is represented by $\gamma_c (= \gamma_1 \cap \gamma_2)$, such that $\gamma_u^i, \gamma_\sigma^i, \gamma_c \subset \gamma^i$. The different subsets of the boundaries do not overlap, i.e., $\gamma_\sigma^i \cap \gamma_u^i = \gamma_\sigma^i \cap \gamma_c^i = \gamma_u^i \cap \gamma_c^i = \emptyset$. While both $\gamma_u^i$ and $\gamma_\sigma^i$ represent the known boundary conditions, $\gamma_c$ is an unknown boundary with unknown tractions $t_c^i$ ($t_c^1 = -t_c^2$) which come into effect during any mechanical interaction between the two solids. The contact traction $t_c^i$ at any point in $\gamma_c$ can be decomposed into the normal and tangential parts, $t_N^i$ and $t_T^i$, as

$$t_c^i = (t_c^i \cdot n^i) n^i + (I - n^i \otimes n^i) t_c^i \quad (2)$$

where the first term denotes $t_N^i$, the second term denotes $t_T^i$, and $n^i$ denotes the outward normal at this point with respect to the body $\Omega^i$. The methodology presented in this paper focuses on the finite element decomposition of the pristine domains $\Omega^1$ and $\Omega^2$ using higher-order elements for better capturing the smoothness of the boundary descriptions $\gamma^i$ and evaluating the contact tractions $t_N^i$ and $t_T^i$ acting over $\gamma_h^i$.

### 2.1. Variational formulation

The above-presented strong form of the governing equation is converted into a weak form using variational methods over the domains of solids $\Omega^1$ and $\Omega^2$. For all kinematically admissible deformations $u^i$ contained in the space $\mathscr{U}^i$ and their variations $\delta u^i$ contained in the space $\mathscr{V}^i$, the weak form of the governing equations representing the dynamic equilibrium at any point in time for the solid $\Omega^i$ is

$$\int_{\Omega^i} \rho^i \partial_t^2 u^i \cdot \delta u^i dV + \int_{\Omega^i} \sigma^i(u^i) : \nabla(\delta u)^i dV = \int_{\Omega^i} \rho^i b^i \cdot \delta u^i dV + \int_{\gamma_\sigma^i} t_\sigma^i \cdot \delta u^i d\gamma^i + \int_{\gamma_c^i} t_c^i \cdot \delta u^i d\gamma^i, \quad \forall \delta u^i \in \mathscr{V}^i \quad (3)$$

Here, the spaces $\mathscr{U}^i$ and $\mathscr{V}^i$ are defined as

$$\mathscr{U}^i = \{u^i : \Omega^i \to \mathbb{R}^3 | u^i \in [H^1(\Omega^i)]^{n_{sd}}, u^i = \overline{u}^i \; \forall x \in \gamma_u^i\} \quad (4)$$

$$\mathscr{V}^i = \{\delta u^i : \Omega^i \to \mathbb{R}^3 | \delta u^i \in [H^1(\Omega^i)]^{n_{sd}}, \delta u^i = \mathbf{0} \; \forall x \in \gamma_u^i\} \quad (5)$$

where $H^1(\Omega^i)$ denotes the Sobolev space of functions. The last term in the virtual work eq. 3, representing the virtual work contribution of contact, exhibits the nonlinear nature of contact problems as neither the contact zone $\gamma_c$ nor the contact traction distribution $t_c^i$ are known apriori, and depend upon the evolving states of the two domains.

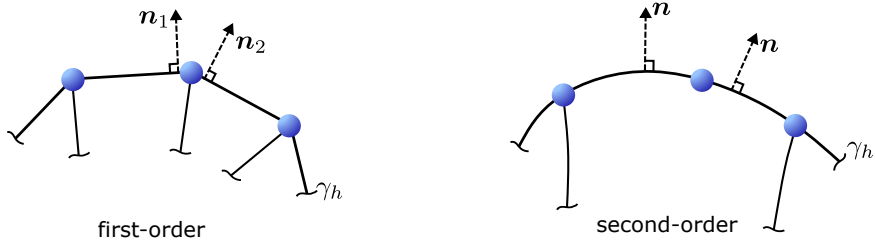

Figure 1: Variation of boundary normal on edges of two first-order elements v/s the edge of one second-order element

For active contact constraint between the solids $\Omega^1$ and $\Omega^2$, the total virtual work of contact is

$$\delta\Pi_c(\boldsymbol{u}, \delta\boldsymbol{u}) = \int_{\gamma_c^1} \boldsymbol{t}_c^1 \cdot \delta\boldsymbol{u}^1 d\gamma + \int_{\gamma_c^2} \boldsymbol{t}_c^2 \cdot \delta\boldsymbol{u}^2 d\gamma \tag{6}$$

$$= \int_{\gamma_c} \boldsymbol{t}_c^1 \cdot (\delta\boldsymbol{u}^1 - \delta\boldsymbol{u}^2) d\gamma \tag{7}$$

Note that if a single body is considered, then as a result of its geometry and boundary conditions, its own boundary might experience a mechanical interaction, thereby leading to a possibility of self-contact. While this virtual work equation is denoted for only two bodies, this can also simultaneously extend to multibody interactions, where contact between any two bodies can be considered similarly.

## 3. Contact problem with higher-order elements

This section presents the unbiased frictional contact interaction between surfaces discretised by higher-order finite elements. Firstly, the geometrical aspects of the contact region construction using a midplane-based methodology are described by taking the second-order Lagrangian elements. This is used to evaluate the normal traction and the normal nodal forces, which are then considered with the unbiased segment-to-segment frictional contact algorithm over midplanes to find the frictional forces between the surfaces. Further, a subsection is devoted to describing the error incurred in the use of decomposed main facets leading to four first-order subfacets and the associated linear shape functions used in lieu of the higher-order shape functions to locate the quadrature points in the parametric space over the midplanes between subfacet-to-subfacet pairs.

### 3.1. Midplane-based contact for higher-order elements

Consider the discretised form of the $i^{th}$ surface $\gamma_h^i$, which is represented by spatial interpolation of the discrete nodal points $\tilde{\mathbf{x}}_j^i$ as

$$\tilde{\boldsymbol{x}}^i = \sum_{j=1}^m \psi_j^i(\xi_1, \xi_2)\tilde{\mathbf{x}}_j^i, \quad \text{where, } \tilde{\boldsymbol{x}}^i, \tilde{\mathbf{x}}_j^i \in \gamma_h^i \tag{8}$$

where $\psi_j^i$ represents the shape function associated with the node $\tilde{\mathbf{x}}_j^i$ and summation is over all total $m$ nodes on the discrete surface. Any contact which is detected by interpenetration between the two surfaces leads to the application of contact interaction between the two surfaces.

The contact interaction is studied through pairwise interaction between discrete facets on the boundaries of the discretised domains. For domains discretised by second-order Lagrangian elements having 27 nodes with the topological connectivity as shown in Fig. 2a, such discrete facets refer to the 9-noded outer facets, which can also be seen as Lagrangian rectangular elements.

In the case of contact between first-order elements as presented in [26], a midplane is constructed between opposite finite element facets by approximating each facet a plane passing through both parametric lines at the parametric centre of the facet. In this work, considering the case of a 27-noded hexahedron element, any outer facet will be a 9-noded quadratic element with a potentially higher curvature than the first-order element. So, while the potential contact pairs are searched using two main 9-noded facets on the opposite surfaces, each main facet is subdivided into four subfacets (see Fig. 2b) and potentially $4 \times 4$ subfacet-to-subfacet pairs are considered for contact interaction.

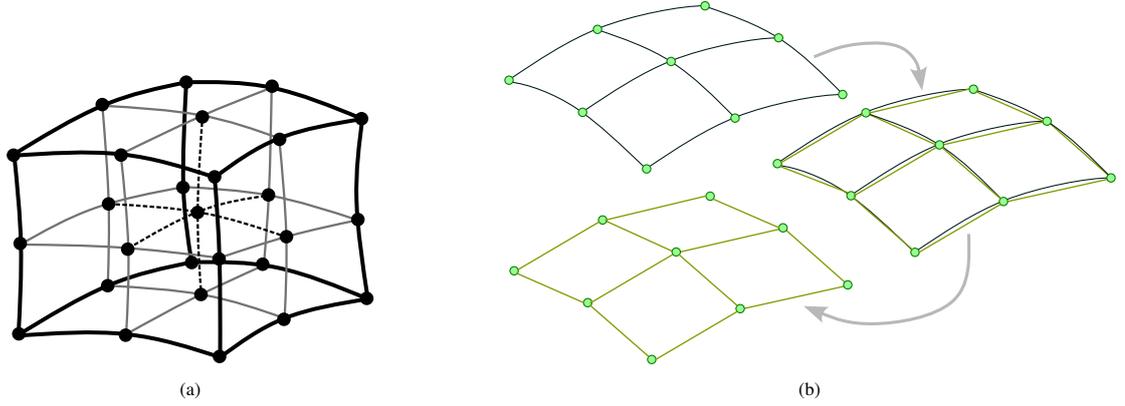

Figure 2: (a) Schematic of a 27-noded hexahedral element, (b) conversion of a curved 9-noded facet of into four bilinear subfacets which are solely used to define respective midplanes in the subfacet-to-subfacet pair.

For any interpenetration between two opposite subfacets, the use of linear penalty results in the generation of normal traction, which is also directly proportional to the interpenetration gap $g_N$ between the two interpolated surfaces, i.e., $\boldsymbol{t}_N = f_s \epsilon_N g_N H(-g_N) \boldsymbol{n}_{\mathrm{mp}}$, where $\boldsymbol{n}_{\mathrm{mp}}$ is normal to the midplane, $\epsilon_N$ is the normal penalty (chosen as minimum of the two bulk moduli), $f_s$ is the scaling factor and $H(\cdot)$ is the Heaviside function. Normal traction, as an effect of interpenetration with respect to the normal plane, is applied equally and oppositely on the two physical facets.

Based on the mathematical nature of the finite element, any surface traction on the outer boundary leads to a set of effective external forces acting on the boundary nodes, the calculation of which depends upon the interpolation functions associated with those boundary nodes. So, using the shape functions on the facet $\gamma_{h_A}$, the set of effective nodal forces $\mathbf{R}$ on this facet can be written as [28]

$$\mathbf{R} = \int_{\gamma_{h_A}} \boldsymbol{\Psi} t \, d\gamma \tag{9}$$

When contact interactions between all pairs of subfacets between main facets are considered, the net force acting on the nodes of the discretised surfaces $\gamma_h^1$ and $\gamma_h^2$ are from overall contribution from all subfacet pairs between them. Note that each subfacet pair has its own midplane, so all the nodal forces due to interpenetration in subfacet pairs are aligned normally to the midplane.

For interpenetration between two 9-noded Lagrangian rectangular facets, the contact region $\gamma_{h_c}$ between two main facets is essentially a union of contact regions between the pairs of subfacets. The subdivision of main facets aids in reducing the inaccuracy in determining the parametric coordinates of each physical facet at the point of intersection of normal passing from the quadrature point with the physical facet. These parametric coordinates are utilised to evaluate the gaps (or heights) of both physical facets from the midplane at quadrature points. A schematic of the contact between main facets using subfacets can be seen in Fig. 3 So, the net nodal forces can be written as the sum of the contribution from each subfacet-to-subfacet pair, i.e.

$$\mathbf{R}^N = \sum_{s=1}^{4\times 4} \int_{\gamma_{h_{c_A}^s}} \boldsymbol{\Psi} t_N^s \, d\gamma^s \tag{10}$$

Here, the normal traction $\boldsymbol{t}_N^s$ is along the direction of the midplane between the respective subfacet-to-subfacet pair, and its magnitude is the penalisation of the normal gap between them. Effectively, this allows us to write the set of

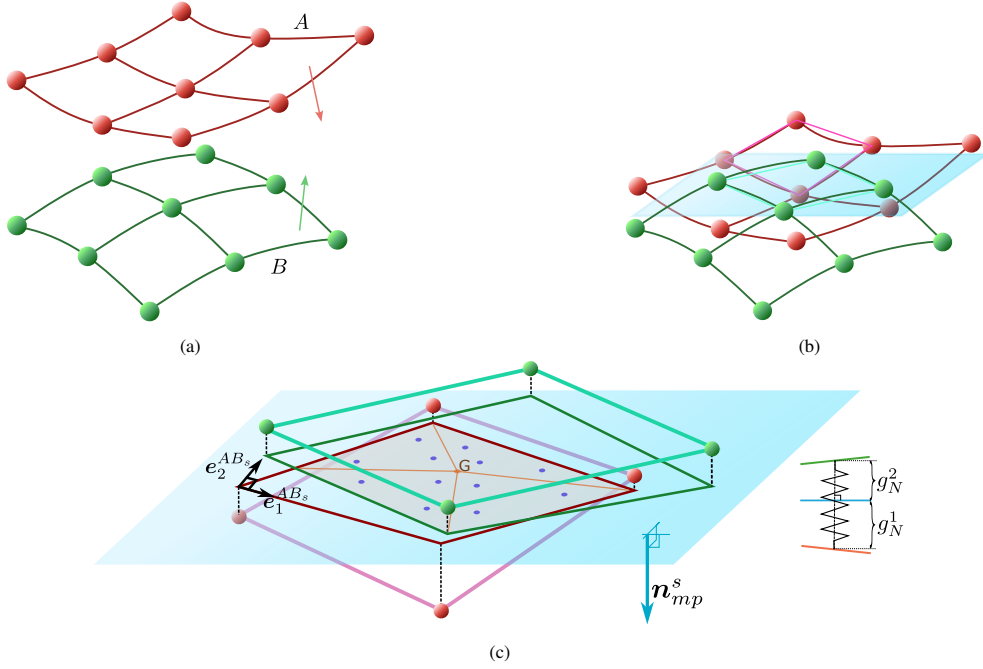

Figure 3: (a) Two 9-noded facets approaching each other (arrows show the outward direction from the underlying 27-noded element) to come into contact at potentially 4 × 4 subfacet-to-subfacet pairs, (b) one highlighted pair of interpenetrating subfacets and the midplane between them, (c) normal projections of the two opposite subfacets over the midplane and the triangulation of the intersection polygon along with the quadrature points in all triangles for traction evaluation. Also shown is the basis for storing frictional traction.

nodal forces on the $i^{th}$ node as

$$\mathbf{R}^N_{i_A} = \sum_{s=1}^{4\times 4} \int_{\gamma_{h^s_{c_A}}} \psi_i \epsilon_N g^s_N H(-g_N) \boldsymbol{n}^s_{\mathrm{mp}} d\gamma^s \tag{11}$$

As these calculations are developed in an average sense, the net contact region between main facet-to-facet pairs is a union of the contact region between subfacet-to-subfacet pairs (defined as the polygon of intersection between their projections over the midplane), so it might not necessarily have the continuity in space. However, the finite element solution developed in this manner is considerably accurate for the full facet-to-facet pair, as demonstrated through multiple numerical experiments.

For progressing towards the numerical solution using quadrature points, the integral over the intersection polygon (contact region) between projections of pairs of subfacets over their respective midplanes can be subdivided over all triangles, and integration is performed using the Gaussian quadrature points over each triangle [29][30]. The interpenetration gap between subfacets is evaluated pointwise by finding the physical points on both surfaces intersecting the virtual line normal to the midplane. The gap of each subfacet at quadrature point $\mathbf{x}_q$ can be written as

$$\boldsymbol{g_N}^i(\mathbf{x}_q) = \sum_{j=1}^{w^i} \psi^i_j \Big|_{(\xi^i_1,\xi^i_2)} \mathbf{x}^i_j - \mathbf{x}_q, \text{ where}, \left(\xi^i_1,\xi^i_2\right) : \mathrm{Proj}_{\gamma_{\mathrm{mp}}}\left(\sum_{j=1}^{w^i} \psi^i_j \Big|_{(\xi^i_1,\xi^i_2)} \mathbf{x}^i_j\right) = \mathbf{x}_q, \mathbf{x}_q \in \gamma_{\mathrm{mp}} \tag{12}$$

where $w^i$ is the number of nodes, and $\mathbf{x}^i_j$ represent nodes on either main facet. The gap vector of interpenetration at each quadrature point $\boldsymbol{x}_q$ can then be written as

$$\boldsymbol{g}_N = \boldsymbol{g}^2_N - \boldsymbol{g}^1_N = \sum_{j=1}^{w^2} \psi^2_j(\xi^2_1,\xi^2_2)\mathbf{x}^2_j - \sum_{j=1}^{w^1} \psi^1_j(\xi^1_1,\xi^1_2)\mathbf{x}^1_j \tag{13}$$

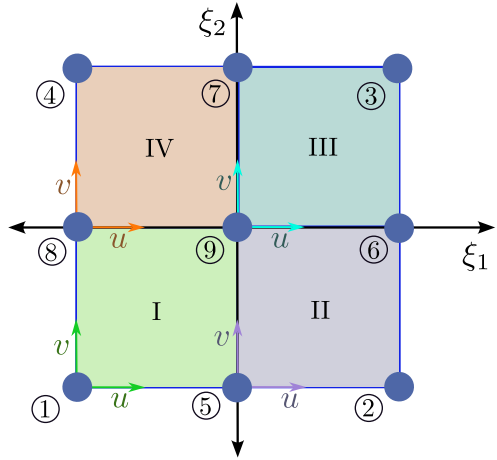

Figure 4: Subdivision in the parametric domain for affine mapping between the main facet and its subfacets in case of a 9-noded Lagrangian rectangular facet

The eq. 12 requires a set of parametric coordinates $(\xi_1^i, \xi_2^i)$ on each $i^{th}$ physical facet. While these can be accurately evaluated using an iterative scheme for the nonlinear shape functions of the main facets, an alternative of using the quadrilateral formed by the projection of linearised subfacets is used here. For each quadrature point on the midplane, a pair of parametric coordinates of bilinear interpolation is found with respect to each projected quadrilateral using inverse bilinear mapping [31]. These parametric coordinates are then mapped to the coordinates $(\xi_1^i, \xi_2^i)$ on each physical facet using an affine mapping, as can be seen in [2][11].

This affine mapping from a projection of linearised subfacet to the main facet is an indirect measure and has been depicted in parametric space in Fig. 4. As an example, the mapping from parametric coordinates $(u, v) \in ([0, 1] \times [0, 1])$ of subfacet I, to the parametric coordinates $(\xi_1, \xi_2) \in ([-1, 1] \times [-1, 1])$ of the main facet is as following

$$u = \xi_1 + 1, v = \xi_2 + 1 \tag{14}$$

The equations of affine mapping for other subfacets with the main facet can be written in a similar manner. Using this affine mapping, the parametric coordinates of physical facets corresponding to the quadrature points can be evaluated. Using the shape functions of the original main facet, the gap (or height) of the physical facet from the quadrature point can be evaluated. This enables the numerical evaluation of the integral for calculating the nodal forces corresponding to the normal traction using the interpenetration gap.

*3.2. Midplane correction*

The construction of the midplane using the subsegments as described previously relies upon the linearised subfacets that do not carry any information about the curvature of the physical surface. This leads to a poor approximation of the midplane. To improve the accuracy of the solution, the midplane can be corrected by utilising the knowledge of interpenetration between the physical surfaces.

The midplane correction fundamentally tries to follow the curvature of the opposite facets corresponding to the centroid of the interpenetration volume. To correct the midplane, an average of all quadrature points weighted by their respective interpenetration gaps is taken to locate the normal projection of the centroid of interpenetration volume over the midplane. As the solution procedure is only affected by the midplane direction and not its position, the midplane can also be assumed to be passing through the centroid itself. Physical points on both surfaces corresponding to this centroid are evaluated. The tangents to the parametric lines at both points are utilised to define the normal directions $\boldsymbol{n}^1$, $\boldsymbol{n}^2$ on the two surfaces. A schematic of this procedure is shown in Fig. 5 for a 2D equivalent problem and a 3D

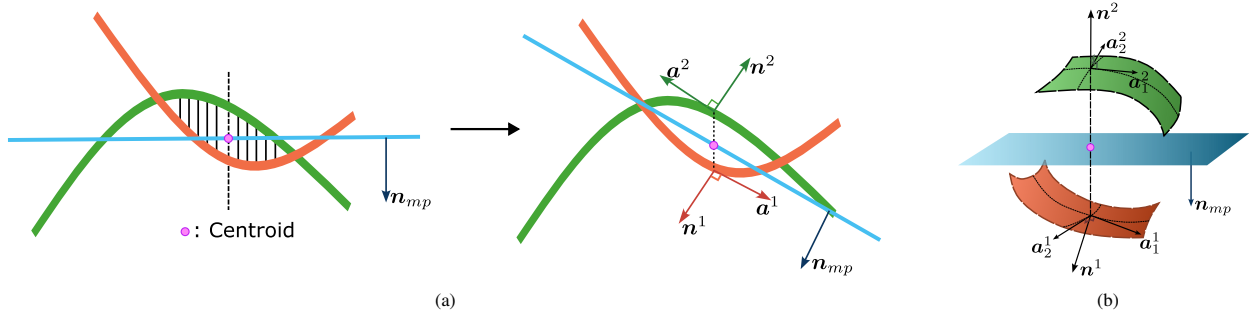

Figure 5: Midplane correction procedure: (a) 2D equivalent showing the initial and final midplane direction, (b) illustration in 3D showing the initial midplane direction and the normal to both physical surfaces corresponding to the interpenetration centroid.

problem. The updated midplane normal can be written as

$$\boldsymbol{n}_{\text{mp}} = \frac{\boldsymbol{n}^1 - \boldsymbol{n}^2}{\|\boldsymbol{n}^1 - \boldsymbol{n}^2\|} \tag{15}$$

This midplane correction can be continued iteratively to further correct the tilt of the midplane, only a single midplane correction has been studied in this work and is influential only for dynamic collision cases.

### 3.3. Unbiased frictional contact

In this work, the frictional contact methodology uses relative sliding between the facets to evaluate the nodal frictional forces in an unbiased manner using the concepts presented in [27]. The frictional effect resisting the tangential motion is also evaluated through the pairwise interaction of subfacets on the interacting boundaries using the predictor-corrector approach. Similar to the use of subfacets for calculating normal nodal forces, the relative motion between the subfacets, about the shared patch of overlap on the midplane, is penalised, and their contribution is added to the nodal forces as frictional forces.

The predicted tangential traction $\mathbf{t}^{\text{trial}}_{T_{n+1}}$ between any pair of subfacets due to relative sliding between solution steps $n$ and $(n + 1)$ causing changes in the local parametric coordinates of the centroid of overlap with respect to opposite subfacets is written as

$$\mathbf{t}^{\text{trial}}_{T_{n+1}} = \mathbf{t}^{\text{up}}_{T_n} + \epsilon_T [(^1\xi^\alpha_{C_{n+1}} - {}^1\xi^\alpha_{C_n})\mathbf{a}^1_{\alpha_{n+1}} - (^2\xi^\alpha_{C_{n+1}} - {}^2\xi^\alpha_{C_n})\mathbf{a}^2_{\alpha_{n+1}}] \tag{16}$$

where $C$ refers to the centroid of overlap and $(^1\xi^1, {}^1\xi^2)$ are the parametric coordinates of the centroid of overlap with respect to the interpolation of one main facet, and $\{\mathbf{a}^1_{\alpha_{n+1}}, \mathbf{a}^2_{\alpha_{n+1}}\}$ denote the co-variant vectors along the parametric lines of both main facets at the centroid. Alternatively, the tangential traction can also be written using the relative velocity of the two facets at the centroid

$$\mathbf{t}^{\text{trial}}_{T_{n+1}} = \mathbf{t}^{\text{up}}_{T_n} + \epsilon_T [(\mathbf{I} - \boldsymbol{n}_{\text{mp}} \otimes \boldsymbol{n}_{\text{mp}})(\dot{\mathbf{x}}^1 - \dot{\mathbf{x}}^2)\Delta t_{n+1}] \tag{17}$$

where $\{\dot{\mathbf{x}}^1, \dot{\mathbf{x}}^2\}$ denote the velocities of the physical point on both main facets corresponding to the centroid of the overlap of both subfacet projections. In the present finite element scheme, this essentially refers to interpolating the nodal velocities of all nodes on the main facets by using their shape functions at the centroid of the overlap of subfacets. This requires evaluating the parametric coordinates $(\xi^1, \xi^2)$ at the centroid for finding the shape functions. While $(\dot{\mathbf{x}}^1 - \dot{\mathbf{x}}^2)$ is the relative velocity in the physical 3D space, its projection over the midplane is considered to define the relative tangential velocity, and subsequently, the displacement which is penalised for predicting the tangential traction.

The predicted tangential traction $\mathbf{t}^{\text{trial}}_T$ for relative motion between any two subfacets is corrected using the return-mapping algorithm in the friction cone for cases where its magnitude exceeds the limiting friction. The corrected (or

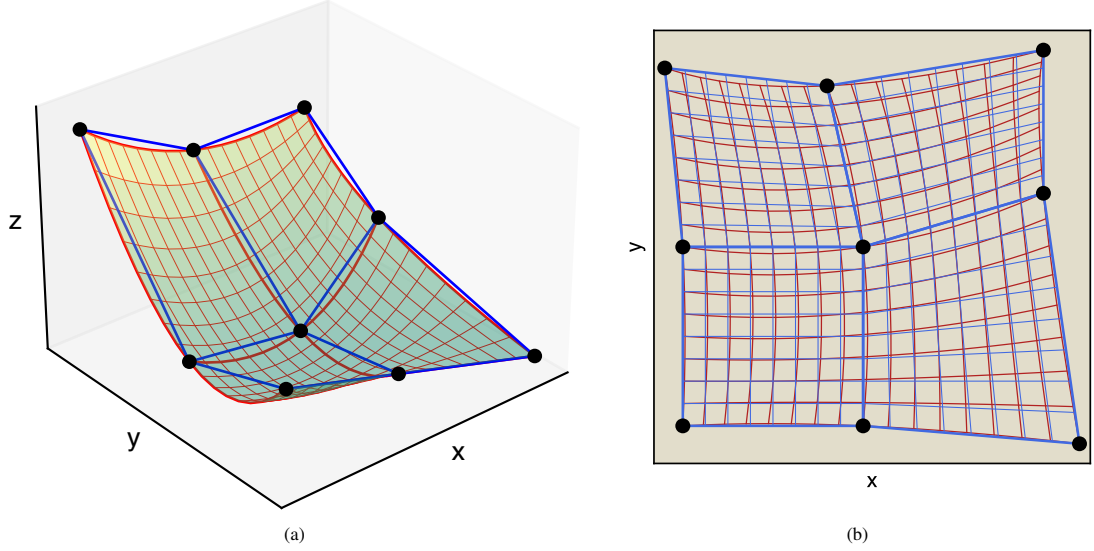

Figure 6: Illustration of the cause of error on using bilinear shape functions for finding parametric coordinates of quadrature points on the midplane. (a) A highly distorted 9-noded Lagrangian shell element with parametric lines (red) and linear edges of subfacets (blue), (b) projection of parametric lines (red) of quadratic element over the z=0 plane and the lines of bilinear interpolation (blue) of subfacets. Notice the differences between the parametric lines of quadratic interpolation and bilinear interpolation on projection in (b).

updated) traction is

$$\mathbf{t}_{T_{n+1}}^{\text{up}} = \begin{cases} \mathbf{t}_{T_{n+1}}^{\text{trial}} & \text{if } \|\mathbf{t}_{T_{n+1}}^{\text{trial}}\| < \mu \|\mathbf{t}_{N_{n+1}}\| \\ \mu \|\mathbf{t}_{N_{n+1}}\| \frac{\mathbf{t}_{T_{n+1}}^{\text{trial}}}{\|\mathbf{t}_{T_{n+1}}^{\text{trial}}\|} & \text{if } \|\mathbf{t}_{T_{n+1}}^{\text{trial}}\| \geq \mu \|\mathbf{t}_{N_{n+1}}\| \end{cases} \quad (18)$$

This tangential traction is applied equally and in opposite directions over the midplane to oppose the relative motion between the two surfaces. To maintain the consistency with the normal nodal forces, the contribution of each subfacet-to-subfacet pair towards the tangential force on the $k^{th}$ ($k = 1, ..., 9$) node in the main facet on one side is written as

$$^{i}\mathbf{R}_k^T = \beta \mu \|^{i}\mathbf{R}_k^N\| \frac{\mathbf{t}_T^i}{\|\mathbf{t}_T^i\|} \quad , \quad 0 \leq \beta \leq 1 \quad (19)$$

$$\beta = \frac{\|\mathbf{t}_T^{\text{up}}\|}{\mu \frac{\|\Sigma_k{}^{i}\mathbf{R}_k^N\|}{A}} \quad (20)$$

where $\mu$ is the coefficient of friction, $\mathbf{R}_k^N$ is the nodal force on the $k^{th}$ node and $\mathbf{t}_T^i$ is the corrected (or updated) tangential traction in the current solution step. It can be observed in the eq. 20 that the predicted-corrected tangential traction provides the direction of frictional forces and indicates the fraction of the limiting friction that acts on the subfacet-to-subfacet pair. It should be noted that the frictional contact between the main facets depends upon the net contribution to the nodal forces by all subfacet-to-subfacet pairs between them, which will all have independent tangential traction calculation.

*3.4. Error of interpolation*

The approximation of subfacets by first-order elements serves two main purposes - (a) defining the initial midplane for each subfacet-to-subfacet pair and projections over these subfacets over the midplane to define the contact region (intersection polygon), and (b) finding the gap between the two physical facets at the quadrature points for contact integral. In the above sections, the affine mapping between the parametric coordinates of bilinear interpolation of subfacets to the parametric coordinates of higher-order interpolation is prone to error due to differences in the straight lines of bilinear interpolation and curved lines of higher-order interpolation. This difference is explained in this section as a potential source of error in the overall contact algorithm that will also help understand the limitations of the similar approach generally seen in the literature [2][11] for higher-order elements.

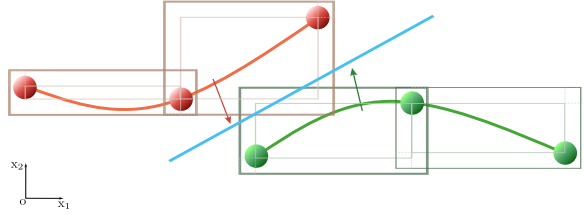

Figure 7: Contact search between two Lagrangian interpolated quadratic facets using bounding boxes around all subfacets.

The physical surfaces formed by first-order and second-order Lagrangian elements in 3D have different mathematical descriptions. While the facets of a first-order element have bilinear interpolation, thereby having an equal spacing between straight parametric lines, the facets of second-order elements have nonlinear interpolation, and the parametric lines are, in general, curved and have unequal spacing between them. Similarly, when the projections of both first-order subfacets and the higher-order main facet over the midplane are considered, the discrepancies in their parametric lines are evident. At each quadrature point in the contact region, the normal gap requires finding the points on opposite main physical facets that normally project over the quadrature point. However, using first-order subfacet and the subsequent affine mapping results in an error when evaluating the parametric coordinates of the required point on the physical facets. This eventually leads to inaccuracies in the contact integral and the evaluated normal nodal forces. The discrepancies in the parametric lines of the projections of higher-order interpolation and the first-order interpolation are illustrated in Fig. 6. Intuitively speaking, if the bilinear interpolation was used on the quadrilateral formed by the four corner nodes of the main facet, there would be an even stronger disparity between its parametric lines and the quadratic parametric lines. It also shows that the four subfacets, with their linear nature, cannot capture the curvature of the main facet at their edges, which also adds to the error in the contact region evaluation.

## 4. Solution algorithm

For computational implementation and numerical solution of any multibody contact problems, the solution schemes generally progress in two stages - spatial search and contact detection. The spatial search is the first stage where intersections between the bounding boxes confining solids are utilised to form a first list of spatial pairs between solids. Using a two-tier approach, the list of potential contacting body pairs is initially formed for the solids with intersecting bounding boxes. Subsequently, a list of potential contacting main facet pairs is formed by running a search algorithm to find main facet pairs with intersecting bounding boxes. This list also includes potential contact pairs on the same body, thereby accounting for any possibility of self-contact. For the sake of simplicity, a case of 2-dimensional (2D) contact between facets with Lagrangian interpolation is shown in the Fig. 7. Similar bounding boxes are considered in a 3-dimensional case (3D), and all possible 4 × 4 subfacet-to-subfacet pairs are tested for interpenetration between them. Using this two-tier contact interaction enables better accounting of the curvature of the interpolated facet and its impact on the evaluated traction.

For contact detection between main facets that confirm interpenetration, the decomposed versions of the main facets in the form of their respective four subfacet are considered to detect contact between subfacet-to-subfacet pairs. Firstly, the intersection between the bounding boxes over the subfacets is considered to reduce the potential subfacet-to-subfacet pairs. The bounding boxes used in this work are axis-aligned and are expanded along each dimension by a small amount to confine the possible curvature effect of the main facets. The subfacet-to-subfacet pairs with intersecting bounding boxes are then run through a series of contact checks to confirm any geometrical interpenetration between them. These geometrical checks over first-order approximation of subfacets progress in the same manner as the interpenetration checks for the contact between solids discretised by first-order elements, as presented in [26] using the clipping algorithm[32] over midplane for defining the contact region for each subfacet-to-subfacet pair.

The subfacet pairs with well-defined contact regions (intersection polygon) are considered for contact traction evaluation using the penalty method based on contact constraint enforcement. Firstly, the normal traction is evaluated for interpenetrating subfacet pairs, and subsequently, the history-dependent frictional traction is calculated for relative motion between the subfacets. The steps involved in these calculations are shown in the Algorithm 1. While the pairs of subfacets are taken into account, the effect of traction over any subfacet results in forces over all nodes due to the mathematical nature of the nonlinear interpolation of higher-order element facets.

One important thing to note is that the distribution of nodal forces, following the nonlinear mathematical description of the finite elements, might result in the direction of nodal forces being outside the facets (i.e. appearing to be opposite to the compressive nature)[4]. These nodal forces, in both their magnitude and direction, are dependent upon the behaviour of all the shape functions of the element that are used in the integral process. These are, however, consistent with the mathematical behaviour of elements to bring out the compressive effect on the displacement of the finite elements. In this regime of nonlinear elements, these nodal forces do not give an intuitive understanding of the traction direction. But, the stress states generated inside the elements are consistent with the traction direction, and their extraction onto the surface is a better representation of the traction.

---

**Algorithm 1** Contact constraint enforcement in normal and tangential direction for each subfacet-to-subfacet pair

---

1: get the polygon of the intersection of projections of opposite subfacets approximated by first-order interpolation
2: triangulate the intersection polygon using edges and its centroid
3: find the area of all triangles
4: set $\left[\Psi_q V_{encl}\right]_i = 0$
5: **for** each triangle **do**
6:     **for** each quadrature point **do**
7:         locate this quadrature point on this triangle using its physical coordinates [29, 30]
8:         find the parametric coordinates $(^i u, ^i v)$ for this quadrature point w.r.t both projected quadrilaterals using inverse bilinear mapping[31]
9:         Use the affine mapping (e.g. eq. 14 to find the corresponding pair of parametric coordinates $(^i\xi_1, ^i\xi_2)$ of higher-interpolation of main facet
10:         find the physical points on both facets at this quadrature point using the parametric coordinates
11:         find the signed gaps $g_N^1$ and $g_N^2$ of both physical facets with respect to this quadrature point on the midplane
12:         find the interpenetration gap $g_N (= g_N^1 + g_N^2)$
13:         **if** $g_N < 0$ **then**
14:             continue to the next quadrature point as there is no interpenetration
15:         **else**
16:             update the quantity $[\Psi V_{encl}]_i$ for both facets by taking contribution from this quadrature point.
17:         **end if**
18:     **end for**
19: **end for**
20: calculate the normal forces $\epsilon_N \left[\Psi_q V_{encl}\right]_i$ on each node of both main facets, with the direction being decided by the midplane
21: find the parametric coordinates and shape functions of higher-order interpolation at the centroid of overlap using the affine mapping
22: find the relative velocity between the centroid using the shape functions of nodes on both main facets
23: using the current time-step increment and the midplane, find the predicted (eq. 17) and corrected tangential traction (eq. 18)
24: calculate the frictional nodal forces using eq. 20

---

## 5. Case of serendipity elements

While this paper focuses on second-order Lagrangian elements for concept description and validation through experiments, the presented methodology can also be extended to serendipity elements by using a virtual node on the 8-noded serendipity quadratic facet to subdivide it into four subfacets. By considering the point in the parametric centre $(\xi_1, \xi_2) = (0, 0)$ of the 8-noded quadratic facet, a physical point on the 8-noded facet can be located ($\mathbf{x} =$ $\sum_{i=1}^{8} \mathbf{x}_i \psi_i(\xi_1 = 0, \xi_2 = 0)$) using the serendipity shape functions. This virtual node will serve as the central node, similar to the case of the 9-noded facet in the Lagrangian element. With such a division, the contact methodology developed for the Lagrangian element can be directly applied here. Here, the similar use of the first-order subfacets will determine the midplane and contact region through the intersection polygon for maximum $4 \times 4$ subfacet-to-subfacet pairs. The main difference will be the change of shape function matrix in the eq. 10, replaced by eight shape functions of the serendipity rectangular element. The use of a virtual node for the purpose of surface subdivision to form the subfacet pairs will not impact the integral as only forces on the physical nodes will be applied, thereby being consistent with the mathematical nature of the facet. This use subdivision is in contrast to the subdivision procedure of [2][11] as it can better capture the local curvature, see Fig. 8a for visualisation of facets with exaggerated distortion having same first eight nodes. The 9-noded facet has just one extra node that changes the surface curvature drastically.

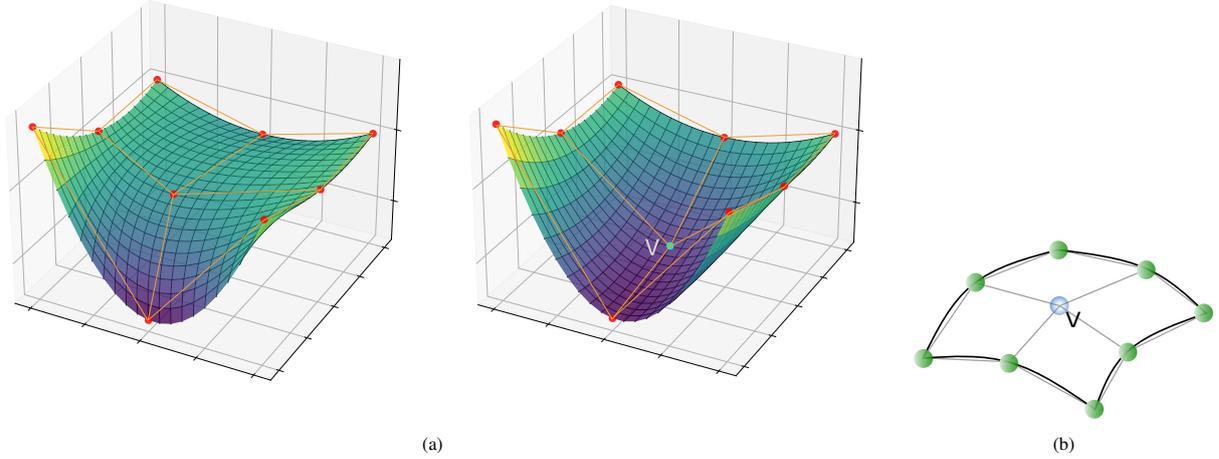

Figure 8: (a) Comparison of subdivision of 9-noded Lagrangian and 8-noded serendipity facet, and (b) schematic for facet subdivision of 8-noded serendipity facet using a virtual node at the parametric centre of the physical facet.

Similar to the case of second-order Lagrangian element, the use of approximating subfacets with first-order interpolation and the mapping between subfacets and physical facets will also induce error in the solution due to the discrepancies in the alignment of the parametric lines of both main facets and subfacets.

## 6. Numerical experiments

The unbiased contact methodology for higher-order elements proposed in this work is studied for its robustness and superiority over first-order elements through multiple examples presented below. Here, the primary objective is to demonstrate the advantage of having higher-order discretisation of the original domain in terms of the accuracy of the solution. The problems presented below include contact interface with flat and curved surfaces, in static and dynamic conditions for small and as well as large deformation problems. Contact patch test, Hertzian contact, and frictional sliding are modelled as quasi-static using the dynamic relaxation technique whereas the remaining cases are modelled as dynamic problems. For the contact integral, 13 Gauss points are used over each triangle to evaluate nodal forces. In this work, paraview [33][34] has been used for visualisation. So, the second-order elements are rendered as linear hexahedra connected by corner nodes, except in the fill rotation problem where true curvature is recovered for clarity in explanation. While the use of linear hex leads to lossy visualisation, the actual physical mesh retains the full geometric accuracy.

### 6.1. Contact patch test

This test checks the robustness of the contact algorithm to transfer uniform contact pressure between two flat surfaces with frictionless contact. Consider two rectangular blocks with the same geometrical size of $20 \times 20 \times 10$

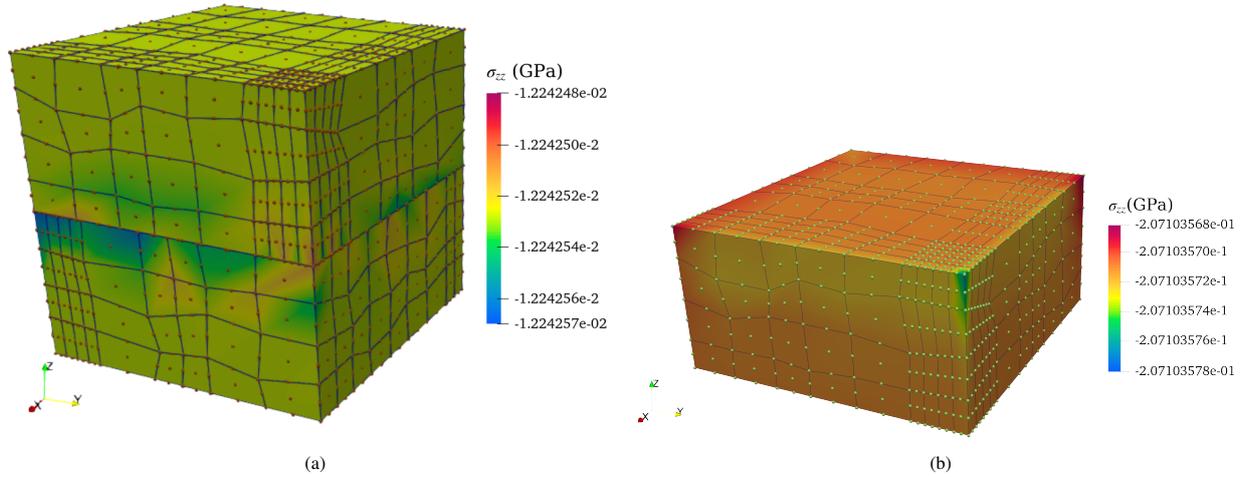

Figure 9: (a) Contact patch test with two blocks arbitrarily discretised using second-order Lagrangian elements, and (b) element patch test by just using the upper block in the contact patch test.

mm$^3$ but different discretisation using 27-noded Lagrangian hex elements with varying sizes to have strong contrast in their discretisation throughout the contact interface; see Fig. 9. The Young's modulus of elasticity and Poisson's ratio for the upper and lower blocks are ($E_1$ = 10 GPa, $\nu_1$ = 0.3) and ($E_2$ = 100 GPa, $\nu_2$ = 0.35), respectively, and a scaling factor of $f_s$ = 10 is used in the contact definition. The top surface of the upper block is pressed downwards by 0.01 mm, while the lateral surfaces of both upper and lower blocks are restricted against motion in their respective normal directions. The distribution of the vertical stress $\sigma_{zz}$ extrapolated to the corner nodes, and the result of the element patch test using just the upper block mesh are also shown in Fig. 9. While theoretically, the same values of stress will be expected throughout the domains in both the contact patch test and the element patch test, the numerical implementation of the finite element scheme introduces machine errors in the solution. However, as can be observed with the digits in the range of stress values, the precision of the contact patch test is nearly the same as the accuracy of the element patch test. So, a direct conclusion can be made that the presented contact algorithm for contact between flat surfaces performs as accurately as the elements underneath the surfaces of contacting bodies.

*6.2. Hertzian contact*

In this classical test based on the Hertzian contact theory[35], two cylindrical bodies are loaded against each other frictionlessly under plane strain assumption. The contact pressure theoretically follows an elliptical (or ellipsoidal in 3D) profile, maximum at the central point and vanishing at the periphery of the contact region. Consider two isotropic elastic bodies as quarter parts of two cylinders with radii $R_1$ = 200 mm and $R_2$ = 250 mm, respectively, and having Young's modulus and Poisson's ratio as $E_1$ = 10 GPa, $\nu_1$ = 0.3, and, $E_2$ = 1000 GPa, $\nu_2$ = 0.35, respectively, see Fig. 10a.

The boundary conditions over the geometries for adhering to the plane strain assumption are as follows: the bottom surface of the lower cylinder is restricted against motion in the vertical direction but allowed to expand horizontally. The surfaces at planes of vertical symmetry and front and back flat surfaces are restricted against motion in their normal directions. The top surface of the upper cylinder is pressed downwards by applying a uniform pressure that corresponds to the case of actual full cylinders having a line loading of P= 9 kN/mm.

Four kinds of discretisations are considered in this experiment, two types each for 8-noded first-order and 27-noded second-order hex elements, such that there are two pairs of first-order–second-order discretisations having the same nodes and just different connectivities for defining the elements, see Fig. 10. In other words, if all elements in each mesh of second-order discretisation are subdivided into eight first-order elements, the corresponding mesh of first-order elements can be obtained. While both first and second-order element meshes have the same number of nodes, the mathematical description of second-order elements will preserve the curvature of the original geometry. In contrast, first-order elements will have linear boundary facets with non-smooth variation of surface normal. This difference in mathematical description is being studied in this experiment to examine the variation in the contact pressure using different types of elements but the same number of nodes, thereby, the same number of degrees of freedom. For strong contrast in non-conformal meshes at the contact zone between the two cylinders, subdomains of both bodies near the interface are discretised with different sizes of elements, as shown in Figs. 10b, 10c, 10d, and 10e.

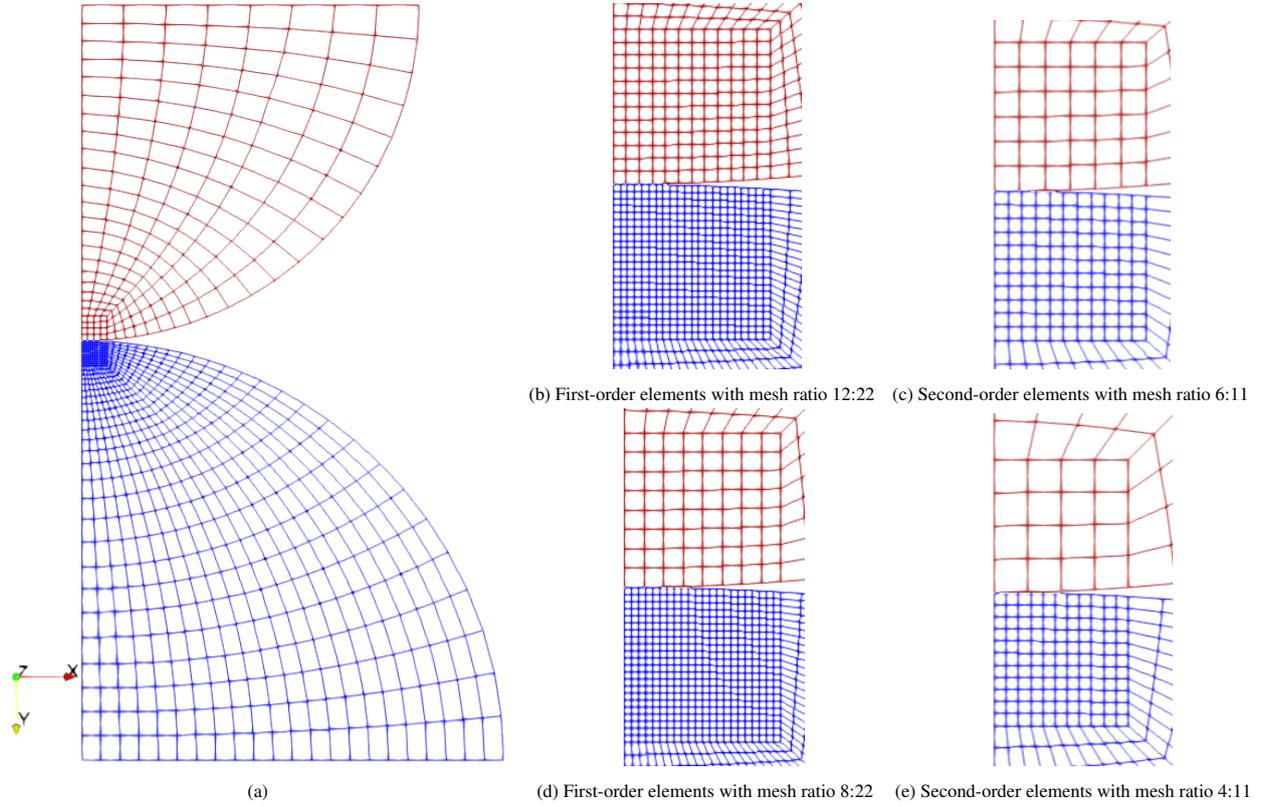

Figure 10: Hertz contact: (a) Discretised cylinders with mesh ratio 4:11 near contact region, (b,c,d,e) show the meshing scheme near the contact region for discretisation with first and second-order elements.

The variation of the contact pressure for both types of elements and discretisation ratios with $f_s = 10$ by directly using the subsurface quadrature points inside elements and their locations are shown in Fig. 11. For first-order elements, the smaller mesh ratio of 12:22 with less non-conformity provides relatively smoother contact pressure variation compared to the higher ratio of 8:22, where the oscillations are significant. This increase in the oscillations is a consequence of the stronger influence of the non-smooth nature of the first-order elements with a higher mesh ratio. On the other hand, the use of second-order elements not only provides a smoother variation of contact pressure for the mesh ratio of 6:11 compared to first-order meshes having the ratio of 12:22, but its use for the higher contrast in mesh ratio of 4:11 is also relatively better compared to its first-order counterpart of mesh ratio 8:22. The increased anomaly in the contact pressure in the outward contact region, on moving from mesh ratio 6:11 to 4:11, is a result of the discrete nature of the finite elements and consideration of a flat midplane for each subfacet-to-subfacet pair. Nevertheless, if the contacting meshes have a higher contrast in the sizes of elements coming into contact, the use of second-order elements provides a better solution compared to the first-order elements. This benefit results from the second-order elements being able to describe the curved physical surface, which gets accounted for in the interpenetration and its subsequent penalisation in the normal traction evaluation. The second-order contact problem shown in Fig. 11d was also tested with the midplane correction; however, the resulting contact pressure variation showed no improvement in this quasi-static test.

For testing effects of softer penalties, the same meshes with second-order discretisation are also studied for $f_s =$

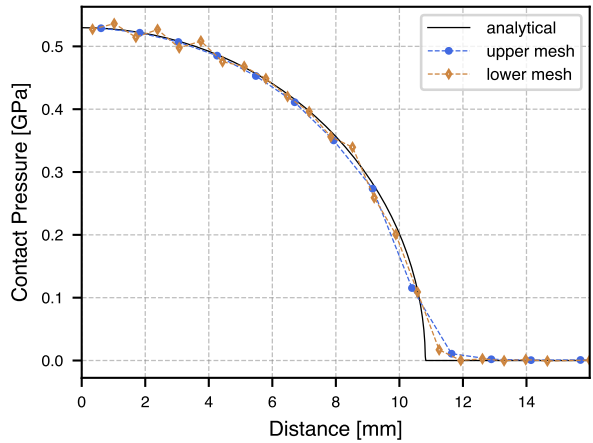
(a) First-order elements with mesh ratio 12:22

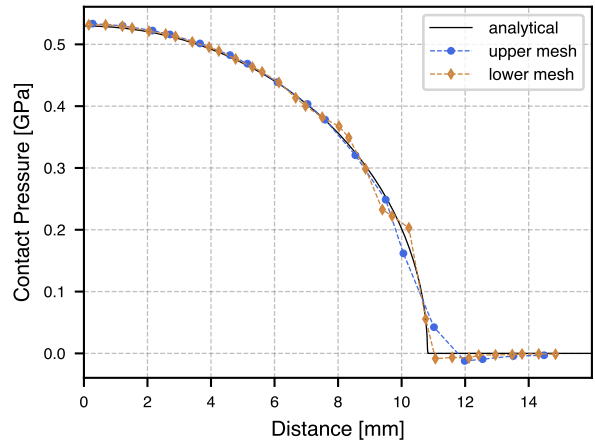
(b) Second-order elements with mesh ratio 6:11

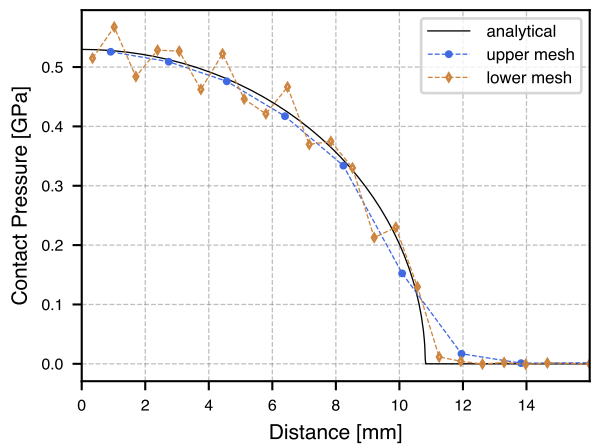
(c) First-order elements with mesh ratio 8:22

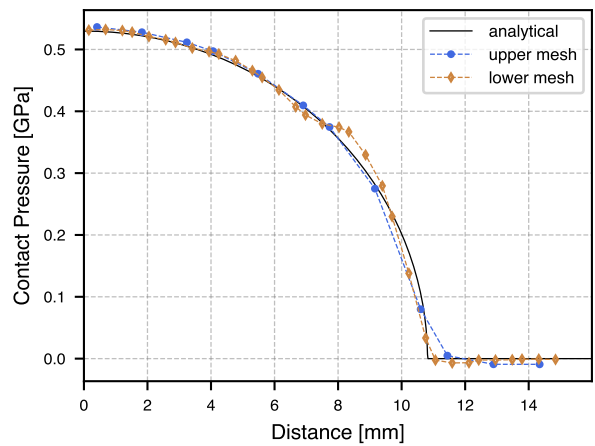
(d) Second-order elements with mesh ratio 4:11

Figure 11: Hertzian Contact: Contact pressure variation between meshes discretised using first-order elements (a, c) and second-order elements (b, d) with both cases having the same number of degrees of freedom in the system and $f_s = 10$ in all cases. Observe that even when the coarser upper mesh is used, the second-order mesh does not have high oscillations like the first-order mesh.

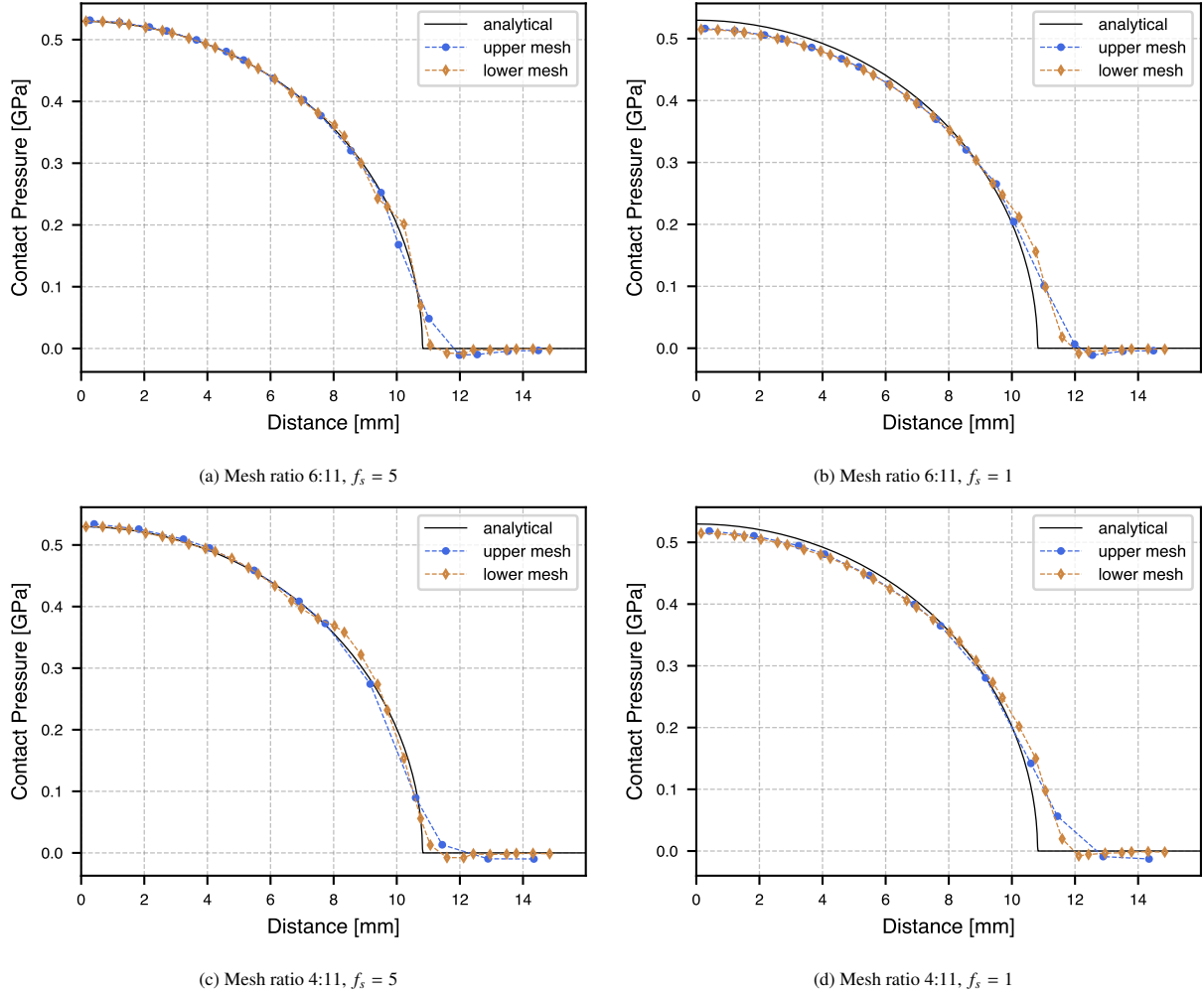

Figure 12: Hertzian Contact: Contact pressure variation for only second-order discretisation for different mesh ratios and penalty scaling factors.

1, 5, and their plots are shown in Fig. 12. As can be seen, using a softer penalty with $f_s = 1$ provides a relatively smoother solution; however, it is at the cost of increased interpenetration, thereby inaccurately predicting a larger contact zone. The smoothness is a result of the increased interpenetration that attenuates the influence of the discrete nature of finite element contact pairs; see the Fig. 13 where mid edge nodes are ignored for simplicity and line segments directly connecting the corner nodes are shown. With softer penalties, the rate of change in the gap along the contact interface between meshes is smaller, so the effect of changes in discrete element features is less pronounced.

### 6.3. Elastic collision of spheres

In this test of frictionless elastic collision between spherical solids, two cases are considered for comparing first-order and second-order Lagrangian elements. In each case, the two types of discretisation schemes have the same nodes, with the only difference being their connectivity for defining the elements. Their different shape functions lead to differences in the physical surface descriptions. So, they both have the same degrees of freedom. All the spherical bodies used in this section have a radius of R = 10 mm, Young's modulus of elasticity, E = 100 GPa, Poisson's ratio $\nu = 0.3$, density $\rho = 0.1$ kg/mm$^3$, and for contact definition, scaling factor is $f_s = 1$. Both cases demonstrate that preserving the original curvature of the surface using second-order elements aids in correctly timed contact detection and subsequent constraint enforcement.

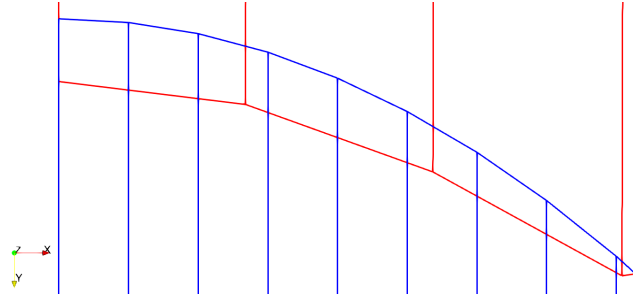

Figure 13: Interpenetration between second-order element mesh with ratio 4:11 and $f_s = 1$, shown by vertically scaling the meshes by a factor of 20 to highlight the gap.

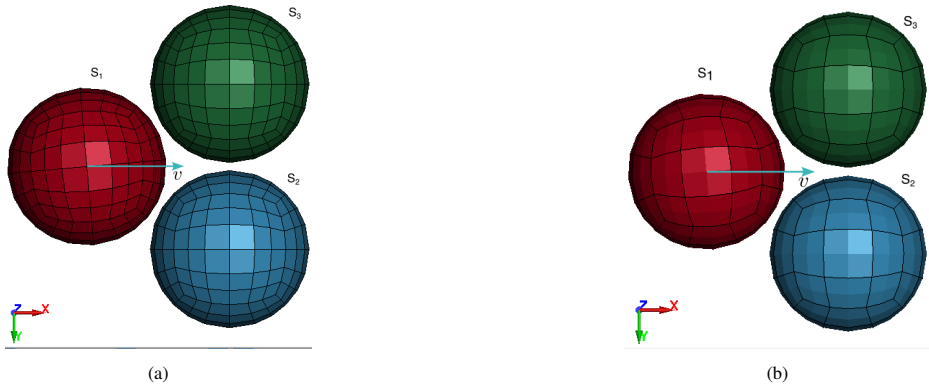

(a)          (b)

Figure 14: Two meshes having same nodes used in collision for (a) first-order elements, and (b) second-order Lagrangian elements

*6.3.1. Ball striking against symmetrically placed balls*

In the first case, three balls are placed in a triangular arrangement, as shown in Fig. 14. While the orientations of the discretised balls $S_2$ and $S_3$ are the same, the ball $S_1$ is rotated by 5° about z-direction to induce a lack of symmetry in the discretised setting of the system. While the physical surfaces in the case of second-order elements will be able to capture the curvature, the case of first-order elements will have a discrepancy with the actual surface and a loss of material in the physical space.

The ball $S_1$ is given a velocity of 0.1 m/s to collide with the two symmetrically placed balls $S_2$ and $S_3$. The time-based changes in momentum in the x and y directions for three different types of discretisations are shown in Fig. 15. Numerous observations can be made here on how the surface description discrepancy between first-order and second-order elements influences changes in momentum. Firstly, the initial momentum of the ball $S_1$ is slightly lower in the case of first-order elements compared to second-order elements as they suffer from material loss due to their straight edges being unable to capture the curvature. The collision in the case of first-order elements is also slightly delayed due to the loss of surface material, hence delayed detection, as can be seen in all plots in $p_x$ and $p_y$. For original spherical geometry, the same momentum $p_x$ will be expected for balls $S_2$ and $S_3$; however, due to higher discrepancy from the original geometry, first-order discretisation has a higher discrepancy in $p_x$. Although second-order elements capture the original curvature, the consideration of finite-sized contact pairs with their respective flat midplanes causes a slight difference in $p_x$. Nonetheless, it is very small compared to the first-order elements. The same reasoning can also be extended to explain the differences in momentum $p_y$ in both cases where the balls $S_2$ and $S_3$ gain similar momentum upon collision in the case of second-order elements with a stark contrast to the error in the use of first-order elements.

*6.3.2. Successive collision in an array of balls*

In the second case, inspired by Newton's cradle, ten spherical bodies are arranged with a uniform gap of 0.5 mm between them as shown in Fig. 16a. Each ball is rotated by 5° about the z-axis with respect to the ball on its left. The

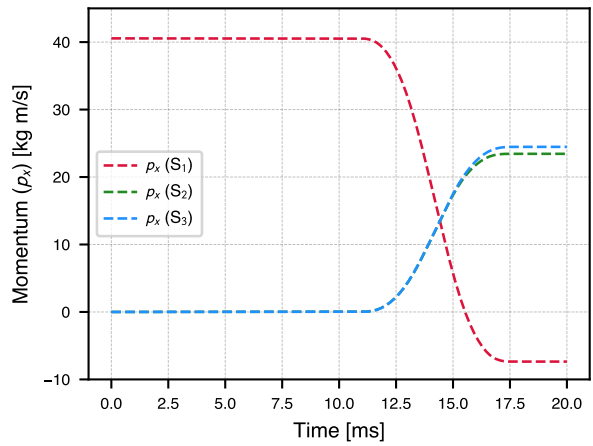
(a)

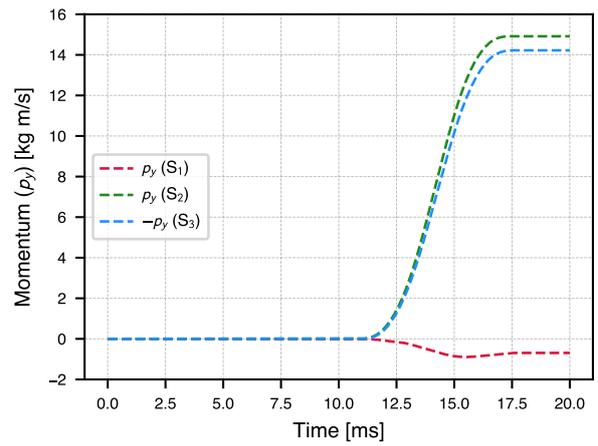
(b)

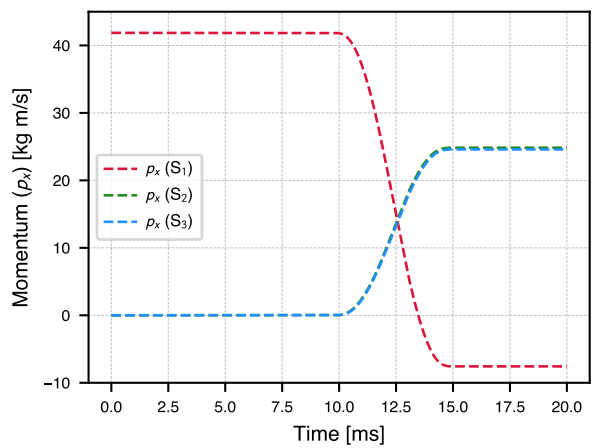
(c)

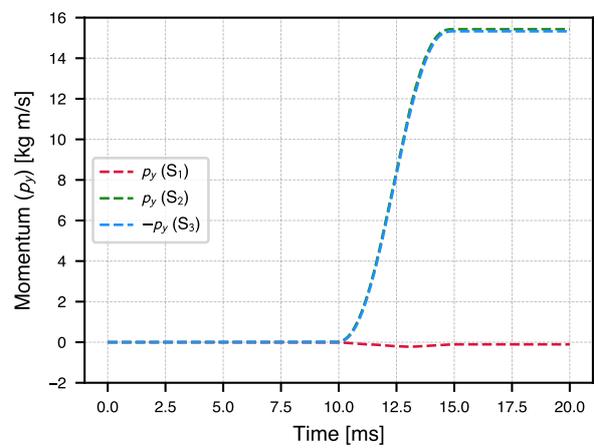
(d)

Figure 15: Symmetric collision of balls: Variation of net momentum of three balls along the x-direction (a,c) and the y-direction (b,d) for first-order elements (a,b) and second-order Lagrangian elements (c,d).

ball $S_1$ is given a velocity of 0.1 m/s and collides head-on with $S_2$, and the motion is transferred to subsequent balls on respective collisions. The changes in the momentum $p_x$ and $p_y$ are shown in Fig. 16 where it can be observed that the momentum $p_x$ is transferred from ball $S_1$ to the last ball $S_{10}$ with successive collisions. The twin benefits of using second-order elements can also be noted here. Firstly, all collisions are quicker in second-order elements with a better representation of curvature, and the error in timing increases until the last collision. Secondly, the error in momentum due to the motion gained in the lateral direction ($p_y$) is much smaller with the use of second-order elements, so it provides superior accuracy.

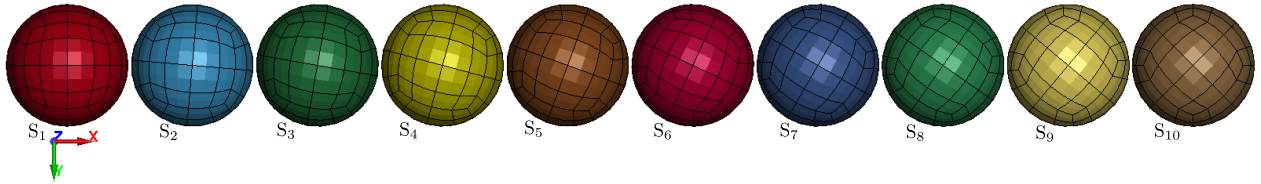

(a)

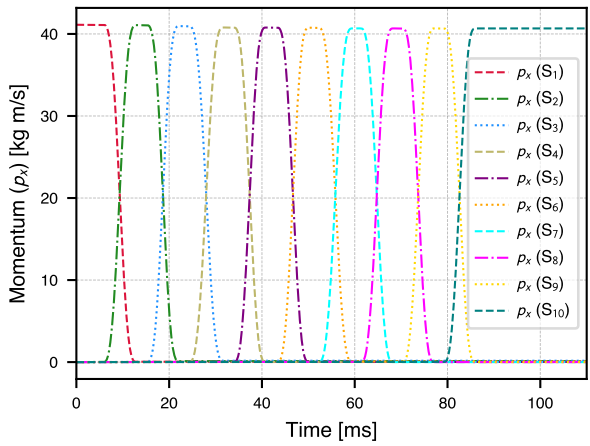

(b)

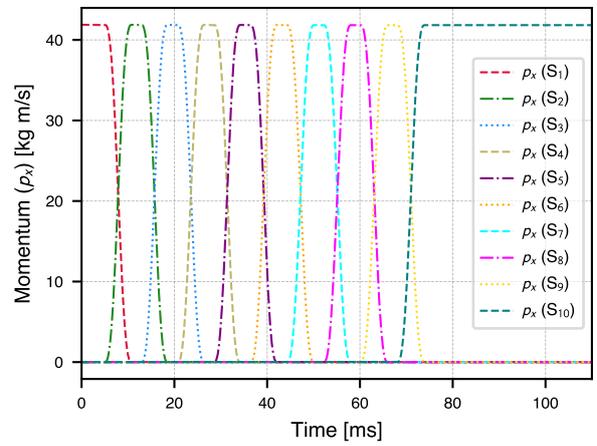

(c)

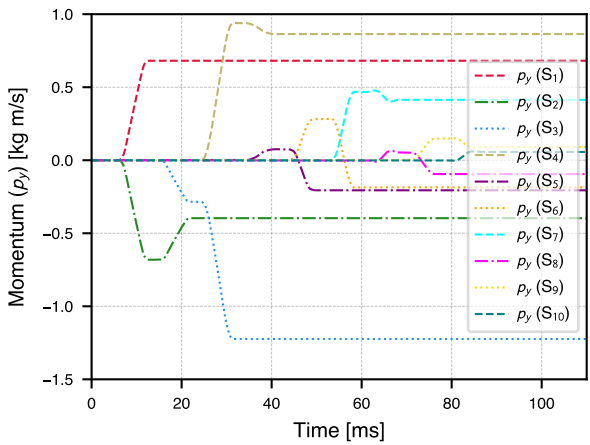

(d)

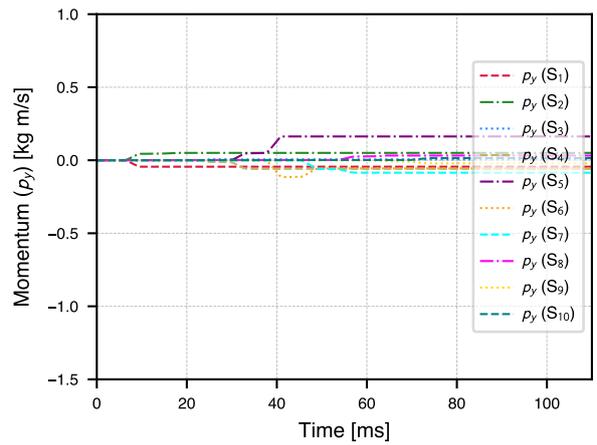

(e)

Figure 16: Successive head-on collision of 10 balls: (a) configuration of spheres, (b,d) momentum variation $p_x$ in first-order elements, and (c,e) momentum variation $p_y$ in second-order elements.

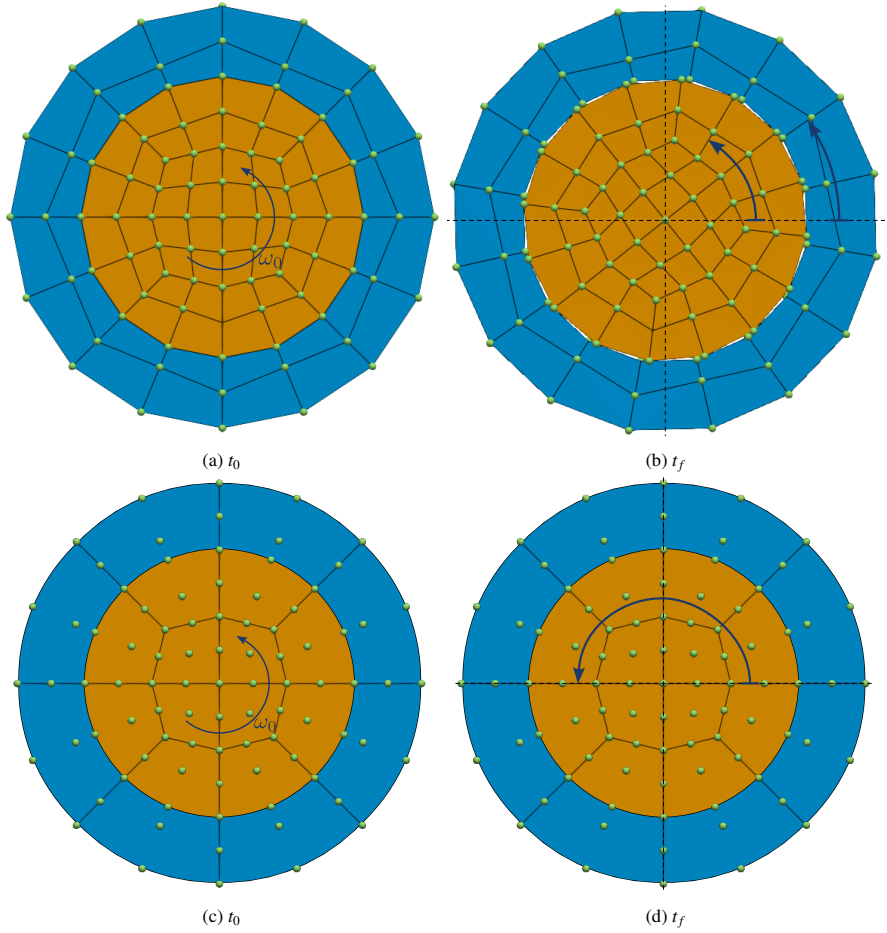

(a) $t_0$  (b) $t_f$

(c) $t_0$  (d) $t_f$

Figure 17: Frictionless rotation of a fill inside a cylinder showing the initial and final configuration for (a,b) first-order discretisation, and (c,d) second-order discretisation. While the cylinder gets rotated with first-order discretisation, it has visibly no rotation with second-order discretisation.

### 6.4. Rotation of a fill inside a cylinder

In this example, taken from [4], the contact algorithm is studied for the frictionless rotation of a fill with a softer material inside a concentric steel cylinder. Both fill and cylinder are composed of isotropic elastic materials with their Young's modulus $E_1$ = 5 GPa, $E_2$ = 207 GPa, Poisson's ratio being $\nu_1$ = 0.4, $\nu_2$ = 0.3, and density $\rho_1 = 7.83 \times 10^{-7}$ kg/mm$^3$, $\rho_2 = 7.83 \times 10^{-6}$ kg/mm$^3$, respectively. The fill has an outer radius of 50 mm whereas the cylinder has an outer radius of 75 mm, and there is a gap of 0.5 $\mu$m at the interface. The fill is given an initial rotational motion such that it can freely cover 180° in 4 ms while the cylinder is initially stationary. Two cases of discretisation are considered, first-order (Fig. 17a, and second-order (Fig. 17c). While the first-order discretisation does not capture the original curvature and has sharp corners, the second-order discretisation perfectly represents the circular curvature with smooth variation of normal along the surface. The non-smooth surface of first-order mesh causes a clattering of fill against the cylinder, leading to slight rotation of the cylinder and the fill loses some angular momentum. On the contrary, the smooth surface variation of second-order mesh results in 180° rotation of the fill while the cylinder remains stationary. Thus, this example perfectly demonstrates the utility of the presented algorithm in cases where representation of curvature is critically important for correct prediction of state changes.

### 6.5. Frictional sliding: Block sliding against a slab

The frictional interaction capability of the proposed contact algorithm is studied here by sliding a deformable block against a rigid slab. The rectangular block of size $5 \times 5 \times 10$ mm$^3$ consists of an isotropic elastic material with Young's

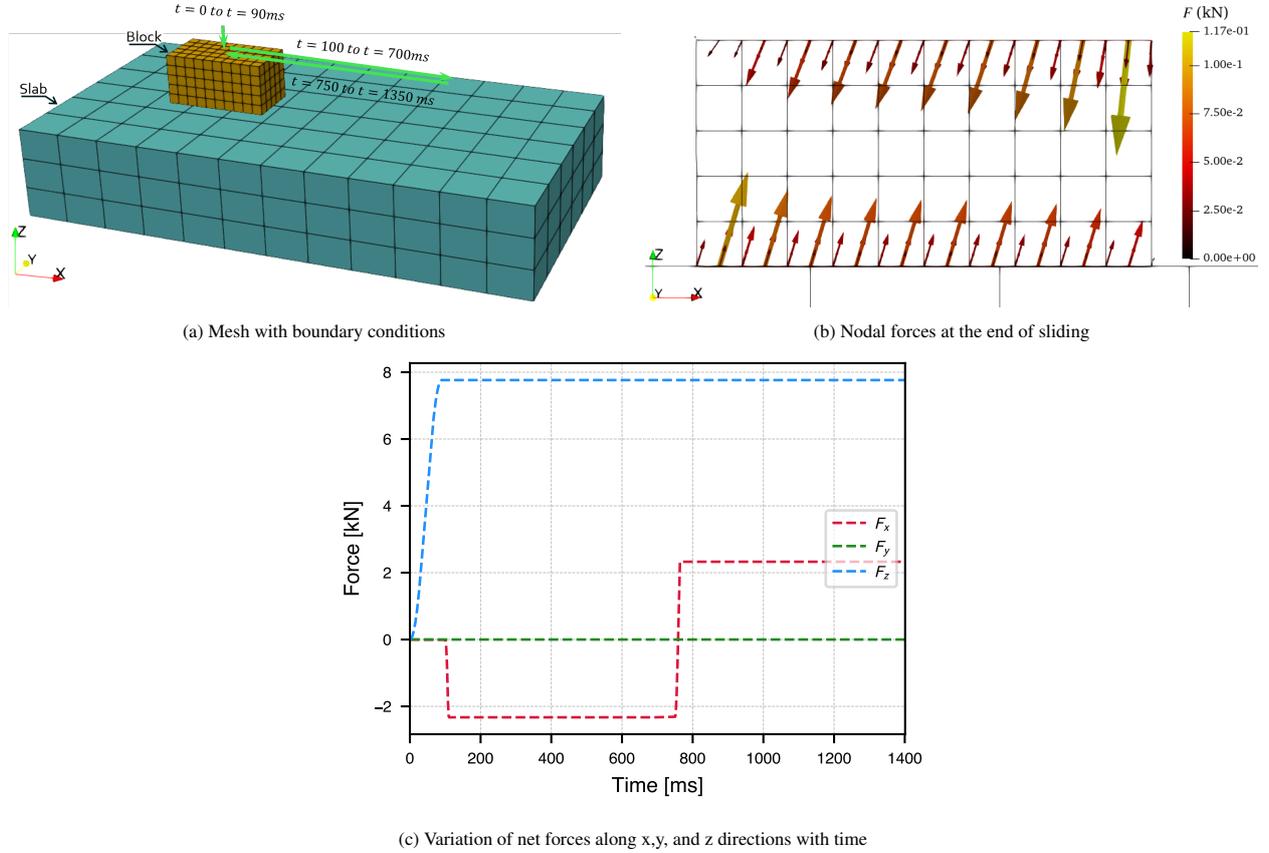

(a) Mesh with boundary conditions

(b) Nodal forces at the end of sliding

(c) Variation of net forces along x,y, and z directions with time

Figure 18: Frictional sliding, with a block being first pressed against a slab and then taken in a to-and-fro motion along it. The retention of frictional force ($F_x$) in (c) at the end of sliding shows the hysteresis effect due to the elasto-plastic analogy of friction.

modulus, E= 207 GPa, Poisson's ratio of, $\nu = 0$, and the contact definition between both bodies has a coefficient of friction, $\mu = 0.3$, and the tangential penalty, $\epsilon_T = 10$. Its top surface is uniformly displaced downwards by $6 \times 10^{-3}$ mm against a rigid slab of size $9 \times 25 \times 50$ mm$^3$ while keeping the scaling factor, $f_s = 1$. It is pushed forward while maintaining the downward push before being returned and held at its original position, thereby completing a full loop, see Fig. 18a. The components of the net force on the contacting surface of the block are shown in Fig. 18c where it can be observed that the normal force ($F_z$) increases during the pressing of the block and stays constant during the sliding. On the other hand, the frictional force ($F_x$) starts increasing in the backward direction during the forward push on the block and stays constant once it reaches the limiting value. As soon as the backward push is initiated on the block, the frictional force, acting backwards, starts diminishing and losing the traction retained from the plastic effect before increasing in the forward direction and staying at the limiting value. Due to the history of dependency on friction and retention of the tangential traction during sliding, the block retains the frictional forces in the contact region at the end of sliding when the top surface of the block is brought to its original position. The nodal forces acting on the top surface of the block to maintain its position and the contact forces that include the frictional effect are shown in Fig. 20. Notice the distribution of the nodal forces follows the mathematical behaviour of traction distribution over second-order Lagrangian elements.

### 6.6. Oblique collision of two rings

In this dynamic test, two identical rings with linear elastic material having Young's modulus E = 0.1 GPa, Poisson's ratio $\nu = 0.49$ and density $\rho = 10^{-6}$ kg/mm$^3$ collide obliquely with equal speed as shown in Fig. 19a. Rings are discretised with second-order Lagrangian elements, and frictionless and frictional cases ($\mu = 0.5$) are considered here. As can be seen, both rings undergo large deformations due to their very high impact speeds. The deformations in the case of frictionless and frictional conditions are shown in Fig. 19. Notice the differences in the configurations of the ring near the contact region and the parts of the rings in contact at $t = 1$ ms due to the frictional effect.

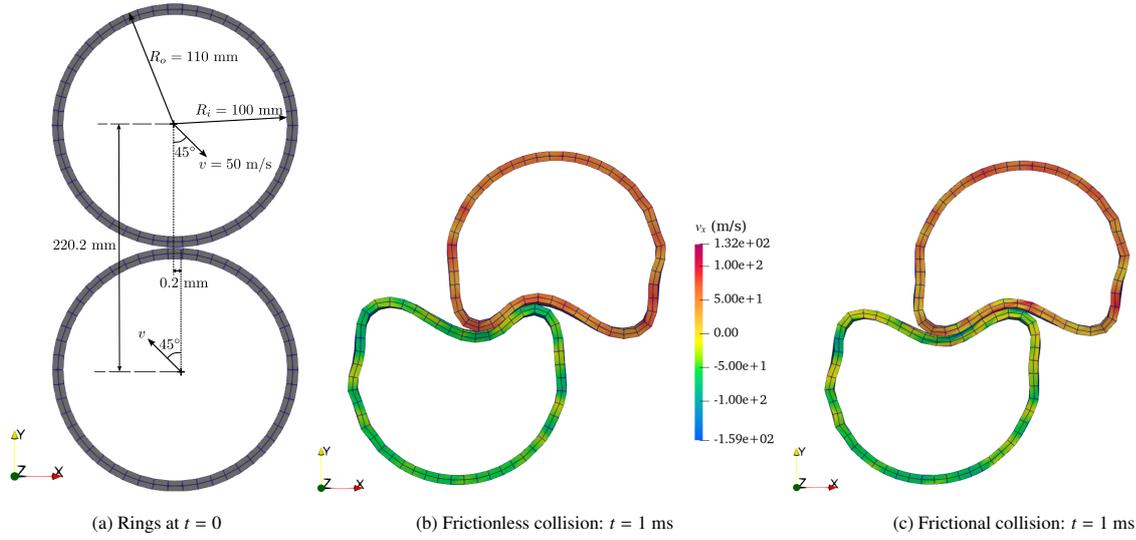

(a) Rings at $t = 0$     (b) Frictionless collision: $t = 1$ ms     (c) Frictional collision: $t = 1$ ms

Figure 19: Oblique collision of identical rings for frictionless and frictional conditions with hex connecting only corner nodes shown for visualisation.

### 6.7. Self-contact - crushing of a tube

The single-pass contact constraint enforcement capability of the presented formulation is of significant importance for self-contact problems. To study this, a problem of axial crushing of a tube is taken in this test. The tube has a length of 50 mm, external and internal diameter of 15 mm and 14 mm, respectively, and is fixed at the bottom. It is perturbed by an inward radial scaling at the bottom by 1% and at 11.11 mm from the bottom by 0.5%. A ring with mass of about 2.28 kg is attached on top of the tube and is relatively rigid, see Fig. 20a. The tube is composed of elasto-plastic material with Young's modulus of elasticity, E= 1.994 GPa, Poisson's ratio $\nu = 0.3$, yield stress $\sigma = 336.6$ MPa, hardening modulus of 1 MPa, and density $\rho = 7.85 \times 10^{-6}$ kg/mm$^3$. Both attached mass and the tube are given an initial downwards velocity of 150 m/s and 75 m/s, respectively. For contact definition, penalty scaling of $f_s = 3.5$ is used in this test. As can be seen in the Fig. 20, the tube undergoes continuous deformation and folds are formed throughout the length of the tube demonstrating self-contact capability of the presented algorithm. While the visualisation using paraview [34] [33] only shows linear hex connecting corner nodes of the second-order element, schematic using quadratic curves on the boundary nodes in Fig. 20d shows smoothness in the deformed shape. This test showing large deformation highlights the utility of the curvature retention capability of second-order elements.

### 6.8. Frictional self-contact

A self-contact test with frictional effect is performed to emphasise the auto-detection of contact between surfaces of a single body along with contact constraint enforcement. This test is studied with both Hex27 and Hex8 elements with the same nodes in the discretised body in the originally undeformed configuration having size $200 \times 200 \times 50$ mm$^3$, see Fig. 21a. The original body discretised with Hex27 elements is shown in Fig. 21b. Its counterpart for first-order elements is obtained by discretising each element in this mesh into eight Hex8 elements, thereby having the same nodes in the two meshes. While the second-order elements have smooth surfaces at all circular surfaces, first-order elements have nonsmooth surfaces at element corners. In both cases, isotropic elastoplastic material with linear hardening is used. It has Young's modulus, E= 10 GPa, Poisson's ratio $\nu = 0.3$, initial yield stress $\sigma_{y_0} = 1.0$, and hardening modulus H= 0.1 GPa. The bottom surface of the block is held fixed while the top surface is prescribed downward displacement by 62.5 mm. For contact definition, the coefficient of friction is $\mu = 0.4$, the normal penalty factor is $\epsilon_N = 10$, and the tangential penalty factor is $\epsilon_T = 10$, and three gauss points per triangle are used in the contact integral. The continuous deformation leads to gradual self-contact between different parts of the surfaces. The deformed configuration for discretisation with Hex27 elements is shown in Fig. 21c overlayed with nodes of both first-order and second-order discretisation. As can be seen, the nodes in the two types of discretisation have position differences around the circular surfaces. This is a result of the differences in first-order and second-order surface representation.

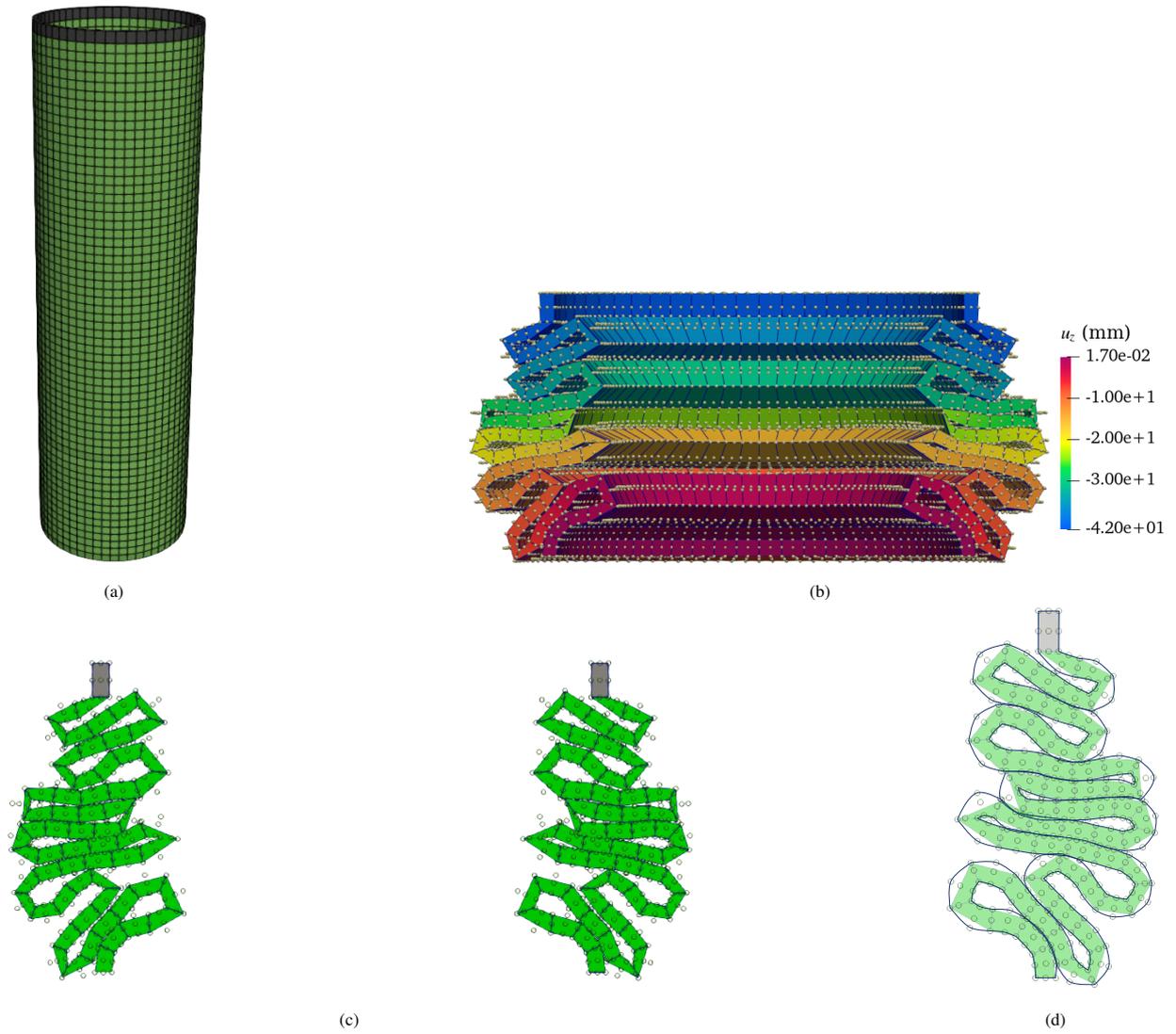

Figure 20: Crushing of a cylinder demonstrating a self-contact scenario: (a) original mesh, (b) deformed configuration of nodes (the solid cells in the visualisation are linear hex connecting the corner nodes of actual second-order elements), (c) a slice section showing the distribution of the nodes, and (d) schematic of the boundary showing smooth curvature.

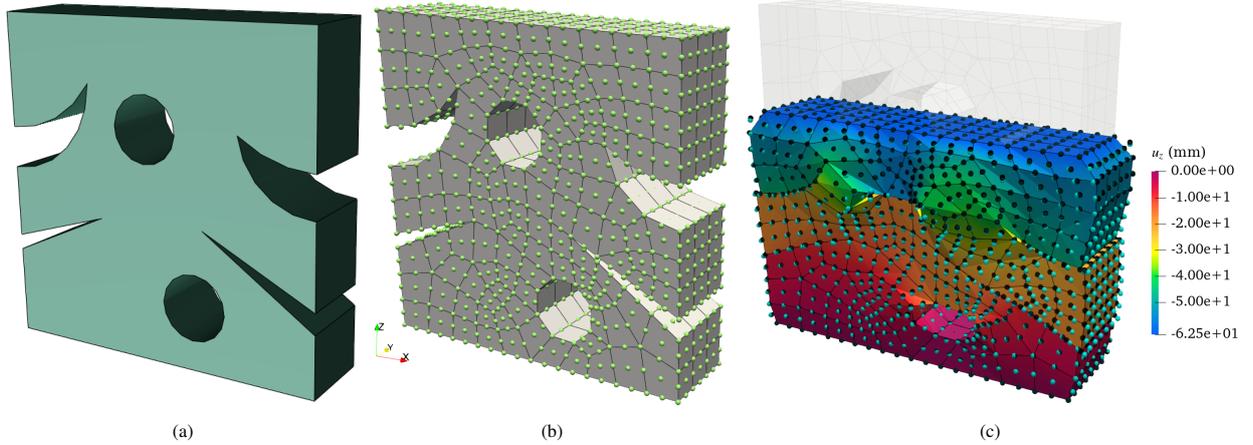

| (a) | (b) | (c) |

Figure 21: Self-contact test: (a) original undeformed geometry, (b) mesh used for Hex27 elements, and (c) deformed mesh along with nodes of both Hex27 (black) and Hex8 (cyan) discretisations. Note that straight edges connecting the corner nodes shown here are a limitation of visualisation, however, the nodal positions are accurately visualised.

*6.9. Inelastic collision of two bars*

In this example of high-rate deformation, two bars with the same original geometry collide axially at the same very high speed, which should ideally result in a flat interface of contact upon deformation as seen in symmetric Taylor's impact test. Here, both first-order and second-order elements are used to discretise the two bars such that they both have the same nodes in the undeformed meshes and the two meshes are non-matching, as shown in Fig. 22. Both bars are made of the same elastoplastic material with linear hardening having Young's modulus E = 69 GPa, Poisson's ratio $\nu$ = 0.33, initial yield stress $\sigma_{y_0}$ = 0.276 GPa, and hardening modulus H = 0.2 GPa. Their maximum dimensions are $10.5 \times 1 \times 1$ mm$^3$, and axial speed is 320 m/s towards each other. Both bars undergo continuous deformation upon collision, with the region near the interface undergoing plastic deformation. The deformed configuration of the two bars is shown in Fig. 23 for both first-order and higher-order discretisations. As can be observed, the deformation in the case of first-order discretisation results in a mismatch in the two bars at the interface, with the right mesh bulging outwards. On the other hand, second-order elements have a relatively flat interface in the contact region upon deformation. Only one side view is shown here to highlight the differences in contact region deformation due to curved surfaces in the original geometry. Also, the overall dimensions of the two deformed meshes are different in both cases. While the two undeformed bars have mismatching element sizes, the second-order elements retain the original curvature, whereas the first-order elements have differences. This difference also results in differences in the overall masses of the two bars, thus adding to the asymmetry in the net momentum exchange.

## 7. Conclusion

The unbiased frictional contact formulation presented in this paper for higher-order elements requires only a single pass and comprehensively outmatches the accuracy of first-order elements in both static and dynamic conditions. Penalty based contact constraints are enforced over respective midplanes of subfacet pairs obtained by subdivision of facet pairs. The additional correction of these midplanes allows to capture the local curvatures of the interacting facets and is critical in gaining higher accuracy in contact traction application. While this work utilises the facet subdivision strategy with affine mapping, it also highlights a potential source of error due to the mismatches in parameteric lines of linear and nonlinear interpolation.

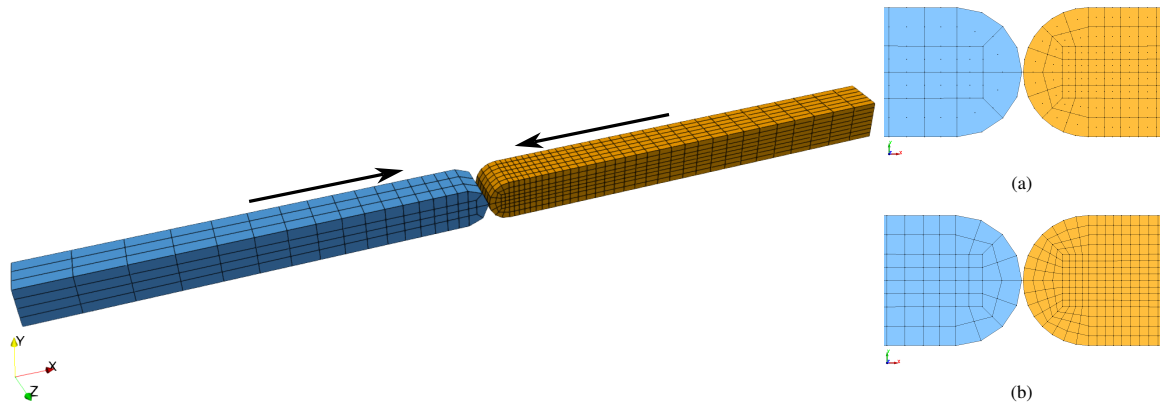

Figure 22: Inelastic collision: (a) two bars with Hex27 discretisation moving towards each other, (b) contacting part of the mesh with Hex27 discretisation, and (c) contacting part of the mesh with Hex8 discretisation. Segmented straight edges in Hex27 mesh are only for visualisation.

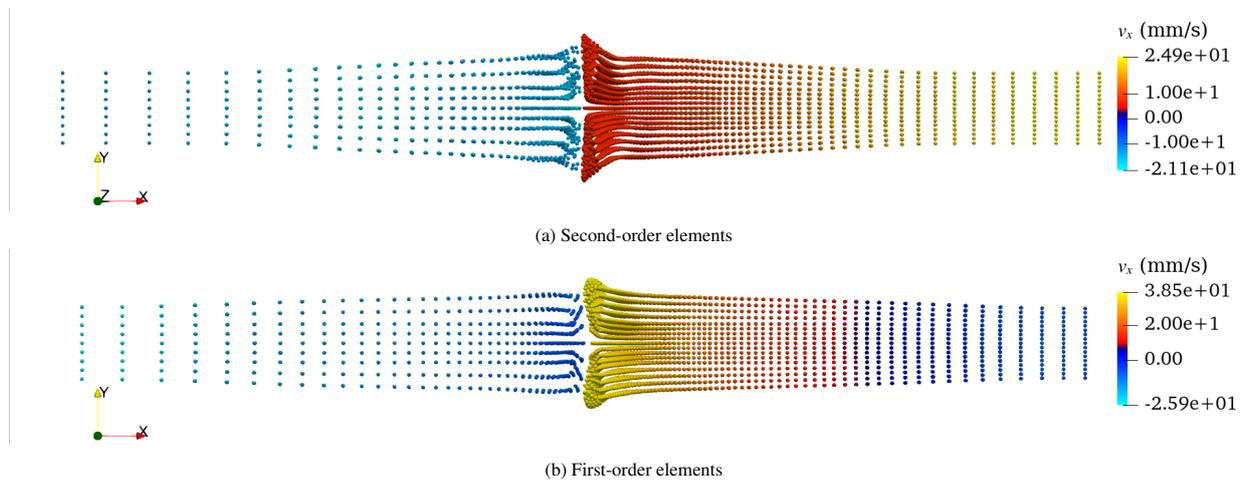

Figure 23: Inelastic collision with dissimilar meshes: Comparison of deformed configurations with only nodes shown for two types of discretisation at time $t = 0.02$ ms.

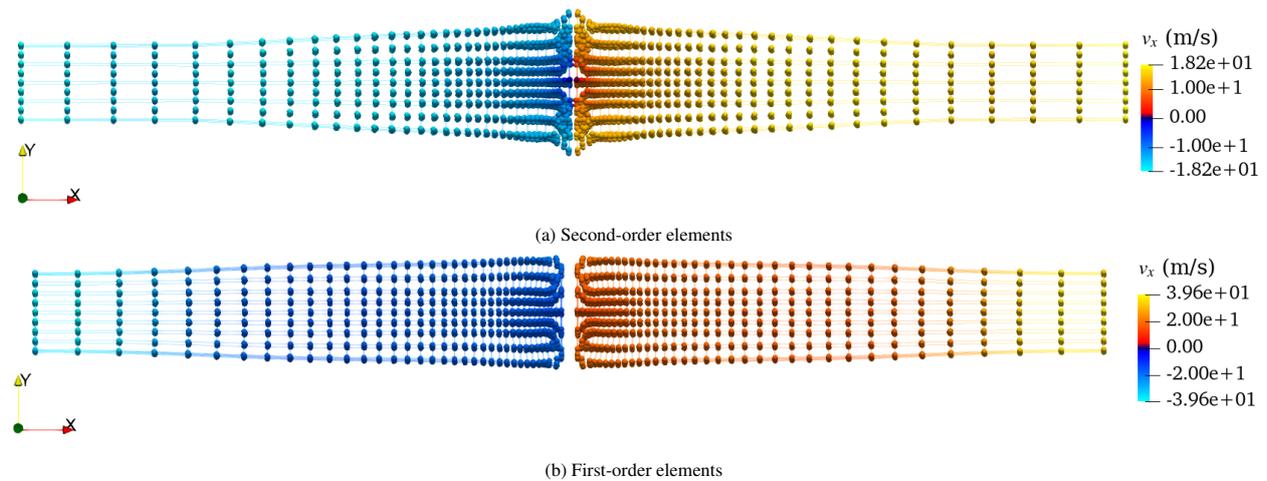

Figure 24: Inelastic collision with similar meshes: Comparison of deformed configurations with only nodes shown for two types of discretisation at time $t = 0.02$ ms.

In practice, the accuracy of presented algorithm matches accuracy of finite elements for contact between flat surfaces. The nonlinear shape functions of second-order discretisation become especially advantageous for contact between curved surfaces with increasingly non-matching meshes as the corresponding first-order discretisation leads to oscillations in contact pressure. In contrast, the second-order elements maintain closeness to the analytical solution with such an increasing mismatch in meshes. Due to the uniform imposition of penalty constraints over the interacting curved surfaces, the solution of the frictionless dynamic problem of rotation of fill inside a concentric cylinder provided solution with higher accuracy for second-order elements.

The second-order meshes are critically important in large deformation contact problems. In large deformation problems, the discrepancies between the original boundary $\gamma$ and the discretised boundary $\gamma_h$ get accentuated resulting in significant differences in the resulting solution. The second-order elements are able to better represent the original boundary $\gamma$ as they leverage nonlinear shape functions, thus providing more accurate solution with smoothness of deformed surfaces. This is the case in self-contact problems as well as the inelastic collision with non-negligible inertia effects. The use of higher-order elements will also be advantageous in multibody systems where the domino effect of error propagation among components due to boundary conditions of contact can be minimised with better representation of discretised surfaces. In cases of contact between higher-order segments having significant deviation between parametric of linear and nonlinear lines of interpolation, error due to use of linearised subfacets for finding gaps will be higher. So a direct inverse problem of iteratively finding physical points normally to the quadrature points will aid in accuracy in such cases. In practical cases where first-order elements are preferred in the bulk, second-order elements can be used over boundaries along with transition elements in between to aid in a more accurate solution. The midplane and patch based conceptual framework presented in this work can also be advantageous in isogeometric analysis with contact.

## 8. Acknowledgement

The authors extend their gratitude to the Engineering and Physical Sciences Research Council (EPSRC) and Rolls-Royce for providing financial support for this research through Strategic Partnership in Computational Science for Advanced Simulation and Modelling of Engineering Systems - ASiMoV (EPSRC Reference EP/S005072/1).